\documentclass[preprint]{article}

\pdfoutput=1

\usepackage{amsmath,amssymb,amsfonts}
\usepackage[pdftex]{graphicx}
\usepackage{times}
\usepackage{url}
\usepackage{cite}  
\usepackage{bm}
\usepackage{algorithm}
\usepackage{algorithmic}
\renewcommand{\algorithmiccomment}[1]{\bgroup\hfill/*~#1~*/\egroup}

\usepackage{xcolor}
\usepackage{xspace}

\usepackage{color} 

\usepackage{url}
\usepackage{tikz}
\usetikzlibrary{backgrounds}

\newcommand{\abs}[1]{\lvert#1\rvert}
\newcommand{\norm}[1]{\lVert#1\rVert}

\newcommand{\Mod}[1]{\ (\text{mod}\ #1)}

\providecommand{\diag}{\operatorname*{diag}} 
\providecommand{\trace}{\operatorname*{trace}} 

\newcommand{\bbbr}{{\mathbb{R}}}

\newcommand{\bbbp}{{\mathbb{P}}}
\newcommand{\bbbe}{{\mathbb{E}}}

\makeatletter
\renewcommand\fnum@algorithm{\fname@algorithm~\thealgorithm.}
\makeatother

\renewenvironment{thebibliography}[1]{%
   \begin{oldthebibliography}{#1}%
     \setlength{\itemsep}{-.3ex}%
}%
{%
   \end{oldthebibliography}%
}

\usepackage[pagewise,displaymath,mathlines]{lineno}

\usepackage{array}
\newcolumntype{H}{>{\setbox0=\hbox\bgroup}c<{\egroup}@{}}
\newcommand{\ignore}[1]{}

\begin{document}

\title{Solving large-scale interior eigenvalue problems to investigate the
  vibrational properties of the boson peak regime in amorphous materials}

\author{Giuseppe  Accaputo${}^1$, \hspace*{5mm}
  Peter M. Derlet${}^2$, \hspace*{5mm} Peter
  Arbenz${}^1$\footnote{Corresponding author.  Email address: \texttt{arbenz@inf.ethz.ch}}\\[2mm]
  ${}^1$Computer Science Department, ETH Z\"urich, CH-8092 Z\"urich,
  Switzerland\\
  ${}^2$Condensed Matter Theory Group, Paul Scherrer Institut,
  CH-5234 Villigen, Switzerland}

\date{19 February 2019}

\maketitle








\begin{abstract} 
  Amorphous solids, like metallic glasses, exhibit an excess of low
  frequency vibrational states reflecting the break-up of sound due to the
  strong structural disorder inherent to these materials.  Referred to as
  the boson peak regime of frequencies, how the corresponding eigenmodes
  relate to the underlying atomic-scale disorder remains an active research
  topic.
  In this paper we investigate the use of a polynomial
  filtered eigensolver for the computation and study of low frequency
  eigenmodes of a Hessian matrix located in a specific interval close to
  the boson peak regime.  A distributed-memory parallel implementation of
  a polynomial filtered eigensolver is presented.  Our implementation,
  based on the Trilinos framework, is
  then applied to Hessian matrices of different atomistic bulk metallic
  glass structures derived from molecular dynamics simulations for the
  computation of eigenmodes close to the boson peak.
  In addition, we demonstrate the parallel scalability of our
  implementation on multicore nodes.  
  Our resulting calculations successfully concur with previous atomistic
  results, and additionally demonstrate a broad cross-over of boson peak
  frequencies within which sound is seen to break-up.
  \par
\end{abstract}



\section{Introduction}
\label{sec:intro}

In amorphous materials, such as structural glass, sound waves have a meaning
only within a finite range of wavelengths.  At long wavelengths, the
heterogeneity of the amorphous structure self averages and an elastic
continuum emerges.  In this regime, sound is well defined via a linear
dispersion characterized by a velocity set by the continuum's isotropic
elastic constants.  However, as the wavelength reduces, the structural
heterogeneity of the glass is increasingly probed, resulting in a
broadening of a sound mode's frequency spectrum.  When this broadening
becomes comparable to the frequency of the sound wave (the Ioffe--Regel
limit), sound loses its traditional meaning as a propagating plane wave.  In
this frequency range, the density of vibrational modes allowed by the solid
is enhanced (the boson peak), also suggesting a transition from propagating
to more localized or quasi-localized non-propagating modes.  The precise
nature of this transition (or cross-over behavior), and its relation to
glass structure, remains an active area of research~\cite{xwln:07, scrs:07,
  shta:08, momo:09, deml:12, schi:13, msfr:13, scsr:15, bcjp:16, like:16}.

One avenue in which this phenomenon may be studied is via the molecular
dynamics simulation technique -- a particle trajectory method able to
produce structural glasses with atomic scale resolution.  Indeed computer
generated amorphous structures may be generated in which the force on each
atom is identically zero.  The harmonic vibrational properties of this
stable structure, which is at a local minimum of the cohesive energy
landscape, may then be investigated through the local quadratic curvature of
this energy surface.  This is done by realizing that the leading order
deviation in energy of a stable configuration (defined by the $N$ atomic
positions, $\bm{r}_{i}$) may be expressed as the quadratic form
\begin{equation} \label{eq:intro.1}
  E\left(\left\{\bm{r}_{i} + \bm{q}_{i}\right\}\right) =
  E\left(\left\{\bm{r}_{i}\right\}\right) +
  \frac{1}{2}\sum_{i,j=1}^N  \bm{q}_{ij}
  \bm{\Delta}_{ij} \bm{q}_{ij},
\end{equation}
where $\bm{q}_{ij}= \bm{q}_{i}-\bm{q}_{j}$, and $\bm{q}_{i}$ is the
deviation of the $i$th atom from its position $\bm{r}_{i}$.
In the above, $\bm{\Delta}_{ij}$ is therefore the second derivative with
respect to the bond-length deviations $\bm{q}_{ij}$.  $\bm{\Delta}_{ij}$ is
a symmetric $3\times3$ matrix.
The quadratic energy term then defines a linear restoring force for such
deviations, and an equation of motion for the $\left\{\bm{q}_{i}\right\}$
coordinates whose secular form equals
\begin{equation} \label{eq:intro.2}
  \sum_{j=1}^N 
  \left(\delta_{ij} (\omega_{n})^{2} - \bm{H}_{ij}\right)
  \bm{q}_{n,j}=0,
  \qquad 1\le n\le N,
\end{equation} 
where $\bm{H}_{ij}=\bm{\Delta}_{ij}/\sqrt{m_{i}m_{j}}$, $m_{i}$ is the mass
of the $i$th atom, and $\omega_{n}$ is the frequency of the $n$th
vibrational eigenmode $\bm{q}_{n,i}$.  
The energy function, $E\left(\left\{\bm{r}_{i}\right\}\right)$, is usually
determined through an empirical force model which is short range (for
metallic and covalent systems), i.e., which spans a few atomic distances and
results in a sparse matrix $\bm{H}$.

Therefore, in order to calculate the vibrational frequencies,
$\omega_{n}=\sqrt{\lambda_{n}}$, we have to solve a real symmetric eigenvalue
problem
\begin{equation}   \label{eq:intro.3}  
  \bm{H} \bm{q} = \lambda  \bm{q},
  \qquad \bm{H} \in \bbbr^{3N\times3N},
  \quad \bm{q} \in \bbbr^{3N}.
\end{equation}
The regime of frequencies relevant to sound waves and the boson peak regime
are at the lower end of the eigenvalue spectrum.  Early simulation work had
often considered system sizes in the range of several thousand atoms, and
more recent work has considered system sizes up to several hundred thousand
atoms.  Contemporary understanding of the frequency regime of the boson peak
suggests the relevant length scales correspond to those of elastic
heterogeneities -- a length scale which is at least an order of magnitude
larger than an inter-atomic distance.  Thus, if one wishes to study the
transition from well defined propagating sound waves to their break up,
larger system sizes will be needed spanning values of $N$ up to several tens
if not hundreds of millions of atoms.  For such large system sizes, the
boson peak eigenvalue regime is no longer an extremal eigenvalue problem,
since there will now be a significant interval of (lower) eigenvalues
describing the allowed sound waves.  This fact motivates the development of
eigensolver methods which are able to focus on a finite interval of
eigenvalues in the \emph{interior} of the spectrum of $\bm{H}$ and their
eigenvectors.

The shift-and-invert Lanczos (SI-Lanczos) algorithm is the method of choice
for computing interior eigenvalues and corresponding eigenvectors of a
symmetric or Hermitian matrix $\bm{H}$ close to some target~$\tau$.
However, the SI-Lanczos algorithm needs the factorization of
$\bm{H} - \tau \bm{I}$ which is not feasible here for its excessive memory
requirements.  For such cases, the Jacobi--Davidson methods have been
developed~\cite{slvo:96, slvo:00, sles:03}.  To be efficient, they however
need an effective preconditioner to solve the so-called correction equation,
which usually entails its factorization~\cite{ahlt:05}.  In an earlier
study~\cite{scha:15}, we were not able to identify such preconditioners
for~\eqref{eq:intro.3}.

In this work we investigate a technique, known as \emph{spectral filtering},
for solving eigenvalue problems that obviates factorizations
altogether~\cite{siro:96, srvk:96, jksc:99, kngl:13}.  Spectral filtering is
combined in practice with Krylov space methods~\cite{fasa:12, sccs:12} or
subspace iteration~\cite{zstc:06, gkla:17}.  In order for the technique to
be applicable the extremal eigenvalues $\lambda_{\text{min}}$ and
$\lambda_{\text{max}}$ of $\bm{H}$, or, at least, some decent bounds
must be available.
%
%
To compute the eigenvalues in the interval
$[\xi, \eta] \subset [\lambda_{\text{min}}, \lambda_{\text{max}}]$, a
polynomial $\rho$ is constructed such that 
$\rho(x) \ge 1$ in $[\xi, \eta]$ and $\abs{\rho(x)} \ll 1$ away from
$[\xi-\varepsilon, \eta+\varepsilon]$.
The desired polynomial $\rho$ could be an approximation of the
characteristic function $\chi_{[\xi,\eta]}$ associated with the interval
$[\xi,\eta]$.  If $\rho(\bm{H})$ multiplies a vector, (most) of the unwanted
eigenvector components are suppressed.  Therefore, $\rho$ is called a
\emph{polynomial filter}.
%
%
The degree of $\rho$ depends on the width of the interval $[\xi,\eta]$, on
the width $\varepsilon$ of the margins, and the strength of the filter.  The
degree increases if $\eta-\xi$ and/or $\varepsilon$ shrink. A consequence is
that increasing parallelism by slicing the interval $[\xi,\eta]$ is not
scalable.  Interval slicing may however be necessary for memory reasons.  In
our experiments we use polynomial degrees as high as
$d=\mathcal{O}(10'000)$.

The numerical experiments consider that part of the spectrum in which sound
is known to break up in a simplified model of an amorphous solid
corresponding to a Hessian $\bm{H}$ of order $4'116'000$ corresponding to
$1'372'000$ atoms.
With the new approach we can deal with models that are more than five times
bigger than those we reported on previously.  Their simulation requires at
least 360 cores on our compute environment in order to store matrices and
vectors in main memory.  The new algorithm relies heavily on matrix-vector
multiplication and therefore scales well to higher core counts.

For the amorphous systems investigated in the present work we find a
spectrum of low frequency modes which are well characterized by sound waves.
However as the frequency of these modes increases, the associated decrease
in sound wavelength results in increased scattering with the underlying
microscopic disorder of the amorphous material until eventually the
vibrational modes have little or no sound-like character.  This is the Boson
peak regime and for the largest system size considered, the transition
appears to have a rather extended frequency range suggesting the Boson peak
frequency and the associated break-up of sound is a broad cross-over rather
than an abrupt transition.

The paper is organized as follows.  In section~\ref{sec:numerics} we review
the technique of polynomial filtering and suggest a filter that should be
useful in connection with subspace iteration.  In section~\ref{sec:impl} we
give some details on how we implemented our eigensolvers with the Trilinos
software framework.  In section~\ref{sec:experiments} we discuss the
numerical experiments that we conducted in a distributed memory computing
environment.  We draw our conclusion and discuss potential future work in
section~\ref{sec:concl}.


\section{Numerical solution procedures}
\label{sec:numerics}

In this section we discuss the restarted Lanczos algorithm and subspace
iteration, both complemented by polynomial filters, to compute interior
eigenvalues of a matrix.  The Lanczos algorithm has been used for this
purpose, e.g., in~\cite{beks:08, lxvy:16}, subspace iteration, e.g.,
in~\cite{gkla:17,zstc:06}.  In subsection~\ref{ssec:SI} we introduce a
polynomial filter that is particularly suited for the use with subspace
iteration.

\subsection{Spectral projector} 



Let $\bm{H}$ be a real symmetric or complex Hermitian matrix of order $n$
and let
\begin{equation}  \label{eq:intro.2}
  \bm{H} = \bm{U \Lambda U}^* = \sum_{i=1}^n \lambda_i \bm{u}_i\bm{u}_i^*,
  \qquad \bm{U} = [\bm{u}_1, \ldots, \bm{u}_n],
  \quad \bm{\Lambda} = \mbox{diag}(\lambda_1,\ldots, \lambda_n),
\end{equation}
with orthogonal/unitary $\bm{U}$, be its spectral decomposition.  For
convenience, we assume that $\bm{H}$'s spectrum
$\sigma(\bm{H})\subset[-1,1]$.  If this is not the case then we can enforce
this property by means of the linear transformation
\begin{equation}  \label{eq:transf.1}
  \bm{H}\ \leftarrow\
  \frac{1}{d}(\bm{H} - c\bm{I}), \qquad
  c = \frac{\lambda_{\text{max}} + \lambda_{\text{min}}}{2},
  \quad
  d = \frac{\lambda_{\text{max}} - \lambda_{\text{min}}}{2}.
\end{equation}
Note that the availability of this transformation implies that we know the
extremal eigenvalues $\lambda_{\text{min}}$ and
$\lambda_{\text{max}}$ of $\bm{H}$ or that we at least have decent
lower and upper bounds, respectively, for them.

To compute the eigenpairs associated with the eigenvalues in a prescribed
interval $[\xi, \eta] \subset [-1, 1]$ it is useful to define the
corresponding spectral projector.  To this end, let
\begin{equation}  \label{eq:intro.33}
  \chi_{[\xi,\eta]}(x) = 
  \begin{cases}
    1, & x\in[\xi,\eta],\\
    0, & \text{otherwise,}
  \end{cases}
\end{equation}
be the \emph{characteristic function} for the closed interval $[\xi,\eta]$.
Then, the \emph{spectral projector}~\cite{saad:11} for the eigenvalues in
$[\xi,\eta]$ is given by
\begin{equation} \label{eq:intro.4}%
  \bm{P}_{[\xi,\eta]} \equiv \chi_{[\xi,\eta]}(\bm{H}) =
  \sum_{i=1}^n \chi_{[\xi,\eta]}(\lambda_i) \bm{u}_i\bm{u}_i^* =
  \sum_{a\le\lambda_i\le b} \bm{u}_i\bm{u}_i^*.
\end{equation}
The orthogonal projector $\bm{P}_{[\xi,\eta]}$ has eigenvalues 0 and 1.  Its
range $\mathcal{R}(\bm{P}_{[\xi,\eta]})$ is spanned by the eigenvectors
$\bm{u}_i$ with $\lambda_i \in [\xi,\eta]$.  The trace of the projector
$\bm{P}_{[\xi,\eta]}$ is the number of eigenvalues in $[\xi,\eta]$, counting
multiplicities,
\begin{equation} \label{eq:intro.5}
  \mu_{[\xi,\eta]} \equiv \trace \bm{P}_{[\xi,\eta]} = \abs{\sigma(\bm{H})
    \cap [\xi,\eta]}.
\end{equation}
%
%

In Algorithm~\ref{alg:ideal} we formulate an \emph{idealized procedure} to
compute the eigenvalues of $\bm{H}$ in $[\xi,\eta]$ and their corresponding
eigenvectors.
\begin{algorithm}
  \caption{Computation of the eigenvectors associated with an interval}
  \label{alg:ideal}
  \begin{algorithmic}[1]
    \REQUIRE Symmetric positive definite Matrix $\bm{H}$ with
    $-1\lessapprox\lambda_{\text{min}}$ and
    $\lambda_{\text{max}}\lessapprox 1$ and an interval
    $[\xi,\eta] \subset [-1,1]$.

    \ENSURE Eigenpairs $(\lambda_1,\bm{u}_1), \ldots, (\lambda_m,\bm{u}_m)$,
    $m=\mu_{[\xi,\eta]}$, with
    $\{ \lambda_1, \ldots, \lambda_m \} = \sigma(\bm{H}) \cap [\xi,\eta]$.

    \medskip

    \STATE Determine an orthonormal basis
    $\bm{V} = [\bm{v}_1, \ldots, \bm{v}_m]$ for
    $\mathcal{R}(\bm{P}_{[\xi,\eta]})$.
  
    \STATE Determine the desired eigenpairs by the Rayleigh--Ritz
    procedure~\cite{parl:80}, i.e., compute the spectral decomposition of
    the (small) matrix $\bm{V}^*\!\bm{HV}$,
    \begin{equation} \label{eq:intro.6}
      \bm{Q}^* (\bm{V}^*\!\bm{HV}) \bm{Q} = \bm{\Lambda}.
    \end{equation}
    The eigenvalues in $[\xi,\eta]$ can now be read from the diagonal of
    $\bm{\Lambda}$; the associated eigenvectors are the respective columns
    of $\bm{U} = \bm{VQ}$.

  \end{algorithmic}
\end{algorithm}
In step~1 of this algorithm it is useful to know (at least an upper bound
of) the dimension $\mu_{[\xi,\eta]}$ of $\mathcal{R}(\bm{P}_{[\xi,\eta]})$.
Then, $\bm{V}$ can be computed by subspace iteration or the Lanczos
algorithm with the matrix $\bm{P}_{[\xi,\eta]} = \chi_{[\xi,\eta]}(\bm{H})$.
Applying $\bm{P}_{[\xi,\eta]}$ to a vector removes all components in the
direction of the unwanted eigenvectors.
%
%
%
%
Algorithm~\ref{alg:ideal} implements an idealized procedure as the spectral
projector $\bm{P}_{[\xi,\eta]}$ is not available.  Forming it would require
the knowledge of the desired eigenvectors.  If $\bm{P}_{[\xi,\eta]}$ was
available the desired subspace could be obtained by one step of subspace
iteration~\cite[Ch.14]{parl:80} provided the subspace is chosen big enough.

In a realistic procedure a function, say $\psi$, is constructed that is much
bigger in $[\xi,\eta]$ than in $[-1,1] \setminus [\xi,\eta]$ such that the
components in undesired directions are suppressed as much as possible
relative to the desired directions if $\psi(\bm{H})$ is applied to a vector.
It is not necessary that $\psi(\bm{H}) \approx \chi_{[\xi,\eta]}(\bm{H})$.

In the sequel we discuss \emph{polynomial filters} $\psi\in\bbbp_d$ that
satisfy
\begin{equation}  \label{eq:filter.6}
  \psi(x) \ge \tau \quad \text{in $[\xi, \eta]$,} \qquad
  \psi(\xi) = \psi(\eta) = \tau, \qquad
  \abs{\psi(x)} \ll \tau\quad \text{in $[\lambda_{\text{min}},a-\varepsilon]
    \cup [b+\varepsilon, \lambda_{\text{max}}]$,}
\end{equation}
with $\tau=\mathcal{O}(1)$ and $\varepsilon\ll1$.
We favor polynomial filters since applying a matrix polynomial to a vector
requires matrix-vector multiplications which can be implemented relatively
easily and efficiently in an HPC environment.
Rational approximations are possible but require the solution of linear
systems which we want to avoid~\cite{bare:16,ytsi:12}.

Note that polynomial filters could in principle be altered during iteration
since all polynomial in $\bm{H}$ have the same eigenvectors.  By changing
the filter different eigenvalues could be exposed.

\subsection{Chebyshev polynomial expansions}

Let $\bbbp_j$ denote the set of polynomials of degree at most $j$.
The Chebyshev polynomials
$T_j(x) = \cos(j\arccos x) \equiv \cos(j \vartheta) \in \bbbp_j$,
$j=0,1,\ldots$, form a complete orthogonal set on the interval $[-1,1]$ with
respect to the inner product
\begin{equation}  \label{eq:cheby-ip.0}
  \langle f,g\rangle\ \equiv\ \int_{-1}^1 \frac{f(x)g(x)}{\sqrt{1-x^2}}\, dx
  = \int_{0}^\pi f(\vartheta) g(\vartheta)\, d\vartheta,
  \qquad x=\cos\vartheta.
\end{equation}
Using the Kronecker delta $\delta_{jk}$, we have
\begin{equation}  \label{eq:cheby-ip.1}
  \langle T_j, T_k \rangle = \frac{\pi}{2}(1+\delta_{0j}) \delta_{jk}.
\end{equation}
Given a piecewise continuous function $f(x)$ defined on $[-1,1]$, we can
form its Chebyshev series
\begin{equation}  \label{eq:cheby-ip.2a}
  \hat{f}(x) = \sum_{j=0}^\infty \gamma_j T_j(x),
  \qquad
  \gamma_j = \frac{\langle f,T_j \rangle}{\langle T_j, T_j
    \rangle}.
\end{equation}
This series converges to $f (x)$ if $f$ is continuous at the point $x$, and
converges to the average of the left- and right-hand limits if $f$ has a
jump discontinuity at $x$.  The polynomial $p\in\bbbp_d$ that best
approximates $f$ in the norm
$\norm{\cdot} = \langle \cdot,\cdot \rangle^{1/2}$ is obtained by
\emph{truncation},
\begin{equation}  \label{eq:cheby-ip.2}
  p_d(x) = \sum_{j=0}^d \gamma_j T_j(x).
\end{equation}

The Chebyshev polynomials satisfy the three-term recurrence
\begin{displaymath}
  T_{k+1}(x) = 2x T_k(x) - T_{k-1}(x),\ k>0, \quad T_0(x) = 1,\ T_1(x) = x.
\end{displaymath}
The three-term recurrence can be conveniently used when a matrix polynomial
is applied to a vector.  Let $\bm{t}_k = T_k(\bm{H}) \bm{x}$.  Then
$\bm{t}_0 = T_0(\bm{H}) \bm{x} = \bm{I}\bm{x}$,
$\bm{t}_1 = T_1(\bm{H}) \bm{x} = \bm{H}\bm{x}$, and
\begin{equation}  \label{eq:3-term-recurrence}
  \bm{t}_{k+1} = 2\bm{H} \bm{t}_k - \bm{t}_{k-1},  \qquad k>0.
\end{equation}
Algorithm~\ref{alg:chebyshev_series} shows how $p_d(\bm{H})\bm{x}$ is
evaluated employing the 3-term recurrence~\eqref{eq:3-term-recurrence}.
\begin{algorithm}
  \caption{Evaluation of truncated Chebyshev series $p_d(\bm{H})\bm{x}$}
  \label{alg:chebyshev_series}
  \begin{algorithmic}[1]
    \REQUIRE Vector $\bm{x}$ and coefficients
    $\gamma_0, \ldots, \gamma_d$ that define $p_d$ in~\eqref{eq:cheby-ip.2}.
    \ENSURE Vector $\bm{y}=p_d(\bm{H})\bm{x}$.

    \smallskip

    \STATE $\bm{t}'' = \bm{x};\ \bm{y} = \gamma_0\bm{t}''$.
    \COMMENT{$\bm{t}'' = \bm{t}_{0}$; $\bm{y} = p_0(\bm{H})\bm{x}$.}
    \STATE \textbf{if} $d\ge1$ \textbf{then} $\bm{t}' = \bm{Hx};\ \bm{y} =
    \bm{y} + \gamma_1\bm{t}'$; \textbf{end if}
    \COMMENT{$\bm{t}' = \bm{t}_{1}$; $\bm{y} = p_1(\bm{H})\bm{x}$.}
    \FOR{$k=2,\ldots,d$}
    \STATE $\bm{t} = 2\bm{Ht}' - \bm{t}'';\ \bm{t}'' = \bm{t}';\ \bm{t}' =
    \bm{t}$. 
    \hfill
    \COMMENT{$\bm{t} = \bm{t}_k$; $\bm{t}' = \bm{t}_{k-1}$; $\bm{t}'' =
      \bm{t}_{k-2}$.}
    \STATE $\bm{y} = \bm{y} + \gamma_k\bm{t}$.
    \COMMENT{$\bm{y} = p_k(\bm{H})\bm{x} = p_{k-1}(\bm{H})\bm{x} +
      \gamma_k\bm{t}_k$.}
    \ENDFOR
  \end{algorithmic}
\end{algorithm}
It constitutes the most time consuming operation in our simulations.  Note
that the degree $d$ can be in the hundreds or even thousands.
Algorithm~\ref{alg:chebyshev_series} presents a stable procedure to
evaluate truncated Chebyshev series~\cite{tref:13}.

\subsection{Dealing with the Gibbs phenomenon}

Truncated Chebyshev series expansions of discontinuous functions exhibit
oscillations near the discontinuities, which are known as \textit{Gibbs
  oscillations} or \textit{Gibbs phenomenon}.  To alleviate this phenomenon
the series must be truncated smoothly using appropriate damping factors.
The damping factors depend on the index at which the series is truncated,
i.e., on the degree of the approximating polynomial.  Thus, $\rho_k(t)$
in~\eqref{eq:cheby-ip.2} is replaced by
\begin{equation} \label{eq:rho_k_smoothened} 
  \rho_k(t) = \sum_{j=0}^k g_j^k\, \gamma_j T_j(t)
\end{equation}
where the $g^k_j$ are the smoothing coefficients.
These coefficients can be determined by different approaches,
see~\cite{wwaf:06} for a survey.

\begin{figure}[bth]
  \begin{center}
    \includegraphics[width=0.66\linewidth]{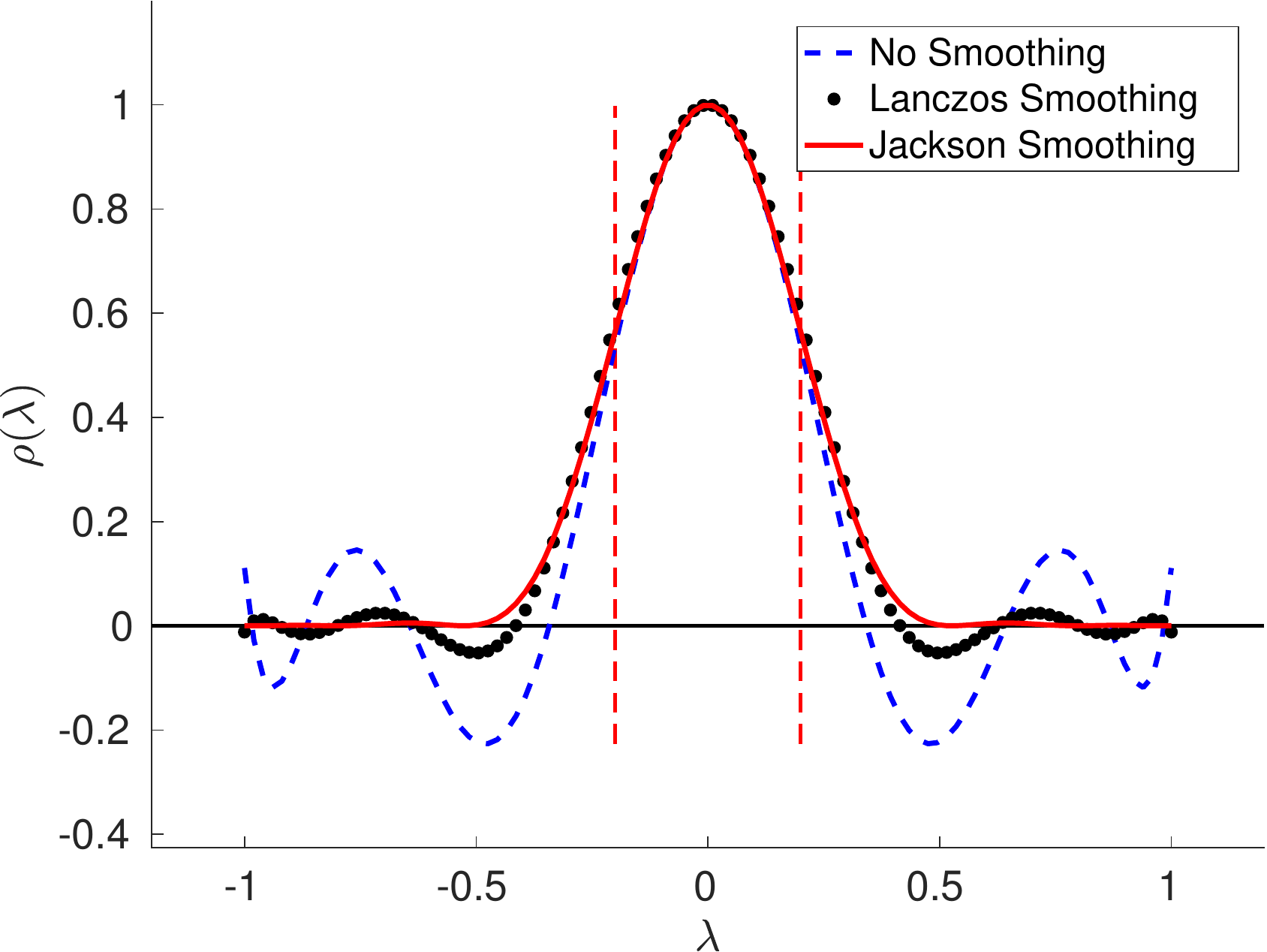}
    \caption{Polynomial filters $\rho_k$ of degree $k = 11$ for the interval
      $[-0.2,0.2]$, using three different smoothing approaches.}
    \label{fig:different_smoothers}
  \end{center}
\end{figure}

Jackson smoothing~\cite{rivl:69, sccs:12} is one of the best known smoothing
procedures, with smoothing coefficients given by
\begin{equation}  \label{eq:jackson_smoothing}
  g^k_j = \left(1 - \frac{j}{k+2}\right) \cos j\alpha_k
  + \frac{1}{k+2} \frac{\cos\alpha_k}{\sin\alpha_k} \sin j\alpha_k,
%
  \qquad \alpha_k =  \frac{\pi}{k + 2}.
\end{equation}
The advantage of Jackson smoothing is its monotonic approximation.  This
implies that the truncated polynomial is positive if the function to be
approximated is so.
%
%
%
Another form of smoothing proposed by Lanczos~\cite{lanc:56}, and referred
to as $\sigma$-damping, uses the damping coefficients
\begin{equation}  \label{eq:lanczos_smoothing}
  \sigma_0^k = 1;\quad
  \sigma_j^k = \frac{\sin j\vartheta_k}{j\vartheta_k},
  \quad j=1,\ldots,k,\quad  \text{with $\vartheta_k = \frac{\pi} {k+1}$}.
\end{equation}
The damping coefficients are small for larger values of $j$, which has the
effect of reducing the oscillations.  The Jackson coefficients have a much
stronger damping effect on these last terms than the Lanczos factors.  In
contrast to Jackson smoothing the Lanczos damping still admits oscillations
and therefore can assume steeper derivatives in the vicinity of jump
discontinuities.
In Figure~\ref{fig:different_smoothers} three polynomial filters
$\rho_{k} \in \bbbp_{11}$ for the interval $[-0.2,0.2]$ are displayed, one of
which is without smoothing and the other two with Jackson smoothing and
Lanczos $\sigma$-damping, respectively.

\subsection{Counting the eigenvalues in an interval}
\label{subsec:counting}

In order that the eigenvalues of $\bm{H}$ in $[\xi,\eta]$ can be computed
numerically their number or at least a (tight) upper bound has to be
known. After all, memory space has to be provided for storing the associated
eigenvectors.
Applying Sylvester's law of inertia for counting the eigenvalues in an
interval is infeasible because of the fill-in generated in the $\bm{LDL}^T$
factorization of $\bm{H}-\lambda\bm{I}$.  However, we showed above that the
number of the eigenvalues in an interval $[\xi,\eta]$ equals the trace of
the spectral projector $\bm{P}_{[\xi,\eta]} = \chi_{[\xi,\eta]}(\bm{H})$
which we do not have available explicitly, but which we can approximate by a
truncated Chebyshev series $\psi(\bm{H})$, i.e.\
$\psi(t) \approx \chi_{[\xi,\eta]}(t)$.
Hutchinson~\cite{hutc:90} showed that
$\bbbe(\bm{v}^*\bm{Hv}) = \trace(\bm{H})$ holds for randomly generated
vectors $\bm{v}$ with entries that are identically independently distributed
(i.i.d.)\ random~variables.
Hutchinson originally used i.i.d.\ Rademacher random variables, where each
entry of $\bm{v}$ assumes the value $-1$ or $1$ with probability $1/2$.  In
general, any sequence of random vectors $\bm{v}_\ell$ whose entries are i.i.d.\
random variables can be used, as long as the mean of their entries is
zero~\cite{beks:08}.  Here, we use a Gaussian estimator to 
approximate $\trace\bm{P}_{[\xi,\eta]}$,
\begin{equation}\label{eq:eigval_count_estimator}
  \mu_{[\xi,\eta]} = 
  \trace\bm{P}_{[\xi,\eta]}\approx 
  \mathcal{T}_M \equiv
  \frac{n}{M} \sum_{\ell=1}^{M}
  \bm{v}_\ell^T \psi_k(\bm{H}) \bm{v}_\ell,
  \qquad \norm{\bm{v}_\ell}_2 = 1,
\end{equation}
by using normally distributed variables for the entries of the random
vectors $\bm{v}_\ell$.  (The factor $n$ in~\eqref{eq:eigval_count_estimator}
is due to the normalization of the $\bm{v}_\ell$.)  Despite the fact that
the Gaussian estimator has a larger variance than Hutchinson's, it shows
better convergence in terms of the number~$M$ of sample
vectors~\cite{avto:11}.  As in~\cite{jksc:99, lisy:16, naps:16} we choose
$\psi$ to be a truncated Chebyshev series for $\chi_{[\xi,\eta]}(t)$, such
that
\begin{displaymath}
  \gamma_j =
  \frac{\langle \chi_{[\xi,\eta]},T_j \rangle}
  {\langle T_j, T_j \rangle},
\end{displaymath}
according to~\eqref{eq:cheby-ip.0}--\eqref{eq:cheby-ip.1}.  For the actual
values of these integrals see~\eqref{eq:app.3} and~\eqref{eq:cheby-ip.1}.




%

\subsection{Computing a basis of
  $\mathcal{R}(\texorpdfstring{\bm{P}}{p}_{[\xi,\eta]})$ with the
  thick-restart Lanczos algorithm}

The desired eigenvectors $\bm{u}_k$ of $\bm{H}$ with eigenvalues
$\lambda_k \in [\xi,\eta]$ span
$\mathcal{R}(\bm{P}_{[\xi,\eta]}) =
\mathcal{R}(\chi_{[\xi,\eta]}(\bm{H}))$.
In Algorithm~\ref{alg:ideal} first a basis for
$\mathcal{R}(\bm{P}_{[\xi,\eta]})$ is computed and then the eigenvectors
are extracted from it by the Rayleigh--Ritz procedure~\cite{parl:80}.
Remember that if 
$\bm{u}_k \in \mathcal{R}(\bm{V})$ then $\lambda_k$ is an eigenvalue of
$\bm{V}^*\!\bm{HV}$.

We consider two different procedures to generate a basis of
$\mathcal{R}(\bm{P}_{[\xi,\eta]})$.  The first is based on the
thick-restart Lanczos algorithm~\cite{wusi:00} where the operator is the
matrix polynomial $\rho_k(\bm{H})$.  Our implementation follows closely the
one described by Li et al.~\cite{lxvy:16} that has been implemented in the
EVSL library%
\footnote{\url{http://www-users.cs.umn.edu/~saad/software/EVSL/}}.
The second procedure is based on subspace iteration with a polynomial filter
designed precisely for this algorithm.  It is presented in the next
subsection.

The requirements for the polynomial filter are different for the Lanczos
algorithm, for the subspace iteration, and for eigenvalue counting.  In the
latter the filter $\psi(t)\in\bbbp_d$ has to be a good approximation of the
characteristic function $\chi_{[\xi,\eta]}(t)$.  As the Lanczos algorithm
converges best towards extremal eigenvalues that are well separated from the
rest of the spectrum~\cite{parl:80}, $\rho_k(t)$ must be (relatively) large
in $[\xi,\eta]$ and small outside.  
%
%
Li et al.~\cite{lxvy:16} suggest, as others before~\cite{wwaf:06, siro:96,
  srvk:96}, to generate a filter that mimicks a Delta distribution, i.e.,
the functional $\delta(\cdot-\gamma)$ defined by
\begin{displaymath}
  \int_\infty^\infty \delta(t-\gamma) \phi(t)\, dt = \phi(\gamma)
\end{displaymath}
for all continuous functions $\phi$.  In the set of polynomials of degree
$k$, $\delta(t-\gamma)$ can be represented by
\begin{equation} \label{eq:cheby-delta.1}%
  \rho_k(x) = \sum_{j=0}^k \frac{T_j(\gamma)}{\langle T_j, T_j \rangle}
  T_j(x).
\end{equation}
$\gamma \in (\xi, \eta)$ is chosen close to the interval midpoint such that
$\tau := \rho_k(\xi) = \rho_k(\eta)$.  By construction, $\rho_k(x) > \tau$
in $(\xi, \eta)$.  
%
Eigenvectors of $\rho_k(\bm{H})$ corresponding to eigenvalues $>\tau$ are
potential eigenvectors of $\bm{H}$.  Care has to be taken, though, as
different eigenvalues of $\bm{H}$ may be mapped to the same value by
$\rho_k$.  Nevertheless, the eigenvectors of $\rho_k(\bm{H})$ corresponding
to eigenvalues $>\tau$ do span $\mathcal{R}(\bm{P}_{[\xi,\eta]})$.  The
correct eigenvalue-eigenvector relations can be found by a Rayleigh--Ritz
procedure applied to $\bm{H}$.  With this filter the eigenvalues close to
$\xi$ and $\eta$ appear usually later than those inside $(\xi,\eta)$.
Since $\mathcal{T}_M$ in~\eqref{eq:eigval_count_estimator} is only an
approximation of $\mu_{[\xi,\eta]}$ we add some 10\% to it to get a
(heuristic) upper bound for the number of eigenvalues in $[\xi,\eta]$.  Of
course, a large overestimation of $\mu_{[\xi,\eta]}$ entails a waste of
memory space.

\subsection{Computing a basis of
  $\mathcal{R}(\texorpdfstring{\bm{P}}{p}_{[\xi,\eta]})$ with subspace
  iteration}
\label{ssec:SI}

\begin{figure}[htb]
  \centering
  \includegraphics[width=0.3\linewidth]{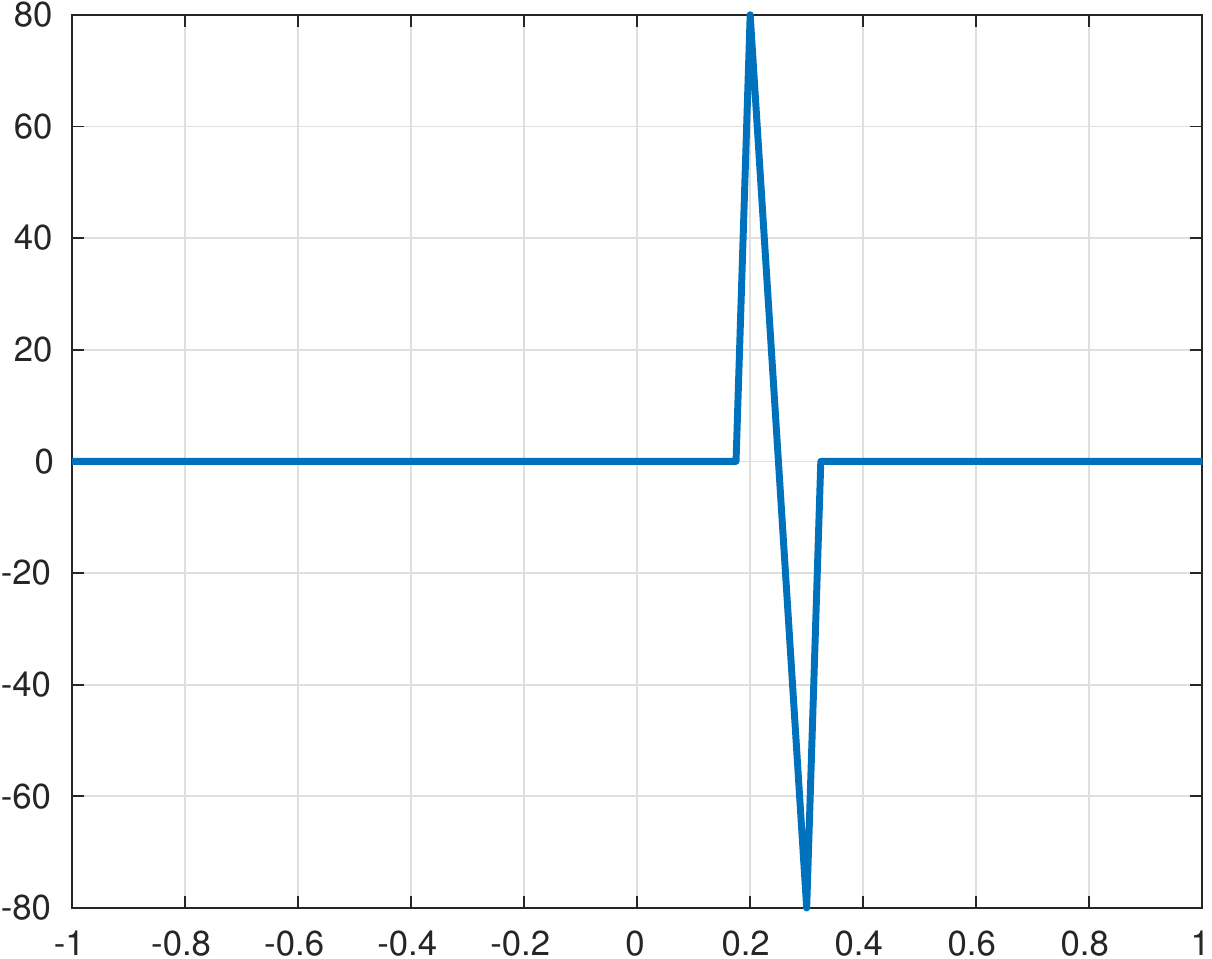}
  \hspace*{0.1\linewidth}
  \includegraphics[width=0.3\linewidth]{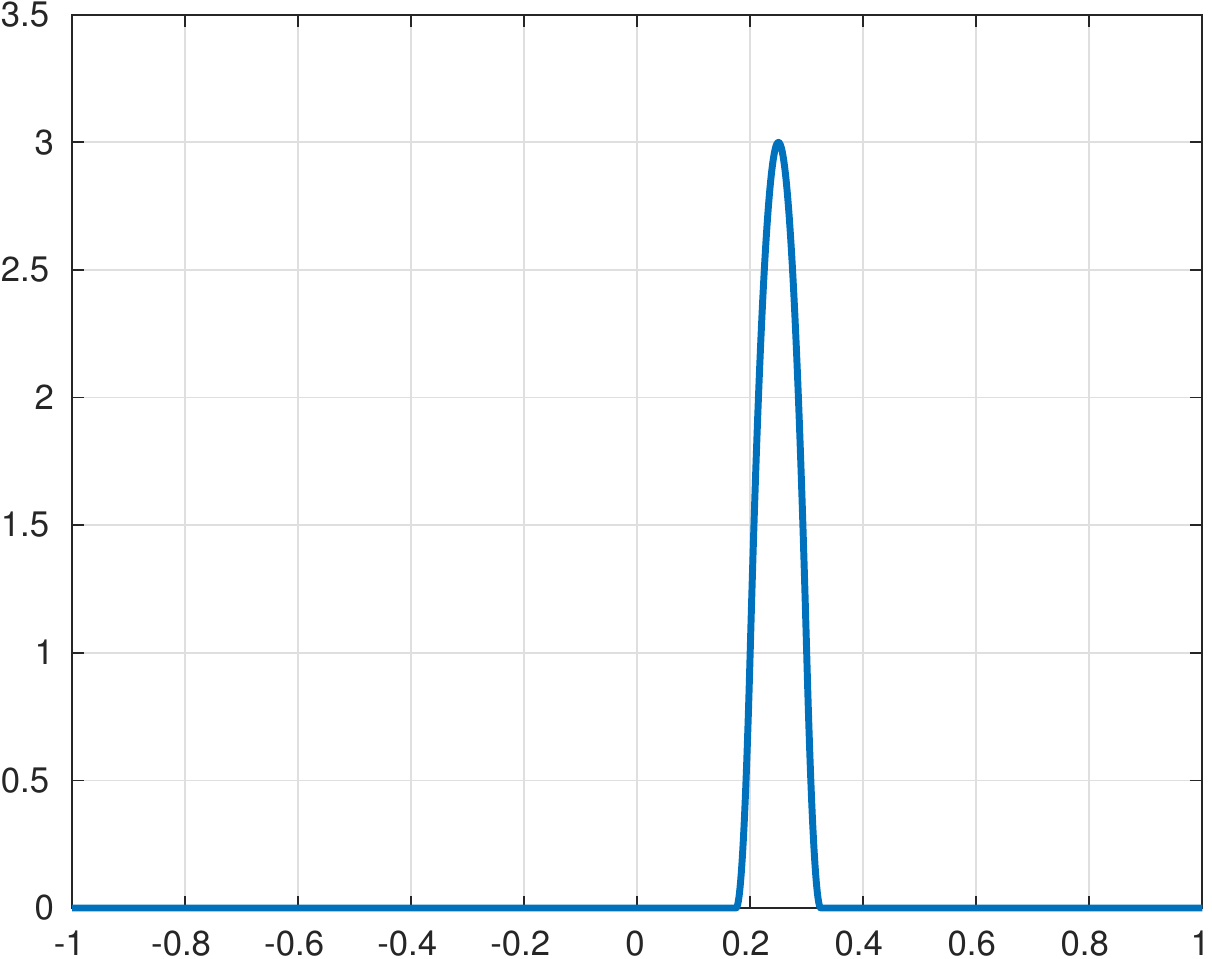}
  \caption{Plot of $q$ (left) and $p$ (right) for $\xi=0.2$, $\eta=0.3$ and
    $\varepsilon=0.025$.  We have
    $q(\xi)=-q(\eta)=\frac{2}{\displaystyle\varepsilon}=80$ and
    $p(\frac{\xi+\eta}{2}) =3$.}
  \label{fig:g}
\end{figure}

In a second approach we compute a basis of
$\mathcal{R}(\bm{P}_{[\xi,\eta]})$ with the subspace iteration.  The
polynomial filter $\pi_k\in\bbbp_k$ is designed to
satisfies~\eqref{eq:filter.6} with $\tau=1$.  $\varepsilon$ should be as
small as possible such that $\abs{\pi_k(t)}$ is smaller than some
prescribed value outside $[\xi-\varepsilon,\eta+\varepsilon]$.  To arrive at
such a filter we first construct a piecewise polynomial $p$ that we then
approximate by a Chebyshev series.  We start with the derivative $q=p'$ that
we define by
\begin{equation}  \label{eq:fct_g}
  q(x) =
  \begin{cases}
    0, & -1 < x < \xi-\varepsilon, \\
    \frac{2}{\varepsilon^2}(x-\xi+\varepsilon), & \xi-\varepsilon \le x < \xi, \\
    \frac{2}{\varepsilon(\xi-\eta)}(2x-\xi-\eta), & \xi \le x < \eta, \\
    \frac{2}{\varepsilon^2}(x-\eta-\varepsilon), & \eta \le x < \eta+\varepsilon, \\
    0, & \xi-\varepsilon < x < 1.
  \end{cases}
\end{equation}
$q$ is a continuous piecewise linear polynomial.  A plot of a typical $q$ is
given on the left of Figure~\ref{fig:g}.

The anti-derivative $p$ of $q$ with $p(\xi)=p(\eta)=1$ is
\begin{equation}  \label{eq:fct_f}
  p(x) =
  \begin{cases}
    0, & -1 < x < \xi-\varepsilon, \\
    \frac{1}{\varepsilon^2}(x-\xi+\varepsilon)^2, & \xi-\varepsilon \le x < \xi, \\
    1 + \frac{1}{2\varepsilon(\eta-\xi)}((\eta-\xi)^2-(2x-\xi-\eta)^2), & \xi \le x < \eta, \\
    \frac{1}{\varepsilon^2}(x-\eta-\varepsilon)^2, & \eta \le x < \eta+\varepsilon, \\
    0, & \xi-\varepsilon < x < 1.
  \end{cases}
\end{equation}
We see that $p(x)>1$ in $(\xi,\eta)$ and that the maximal value that $p$
attains is at the center of $(\xi,\eta)$ where
\begin{displaymath}
  p(\frac{\xi+\eta}{2}) = 1+\frac{\eta-\xi}{2\varepsilon}.
\end{displaymath}
$p$ is symmetric with respect to $(\xi+\eta)/2$, i.e.,
$p((\xi+\eta)/2-x) = p((\xi+\eta)/2+x)$ whenever the values are defined.
The piecewise quadratic $p$ is continuously differentiable by construction.
Therefore, the convergence of the Chebyshev series is quick, the
coefficients $\gamma_j$ in the series decay like $1/j^3$,
see~\cite[Thm.7.1]{tref:13}.  Also the Gibbs oszillations are not as
prominent as with the approximation of a discontinuous function.
If $(\xi,\eta)$ is at or close to the boundary of $(-1, 1)$ we can define
$p(x)$ as above.  In the computation of the coefficients $\gamma_j$ in the
Chebyshev series~\eqref{eq:cheby-ip.2a} only the piece of $p$ in $(-1, 1)$
is to be taken onto account.

\begin{figure}[htb]
  \centering
  \includegraphics[width=0.66\linewidth]{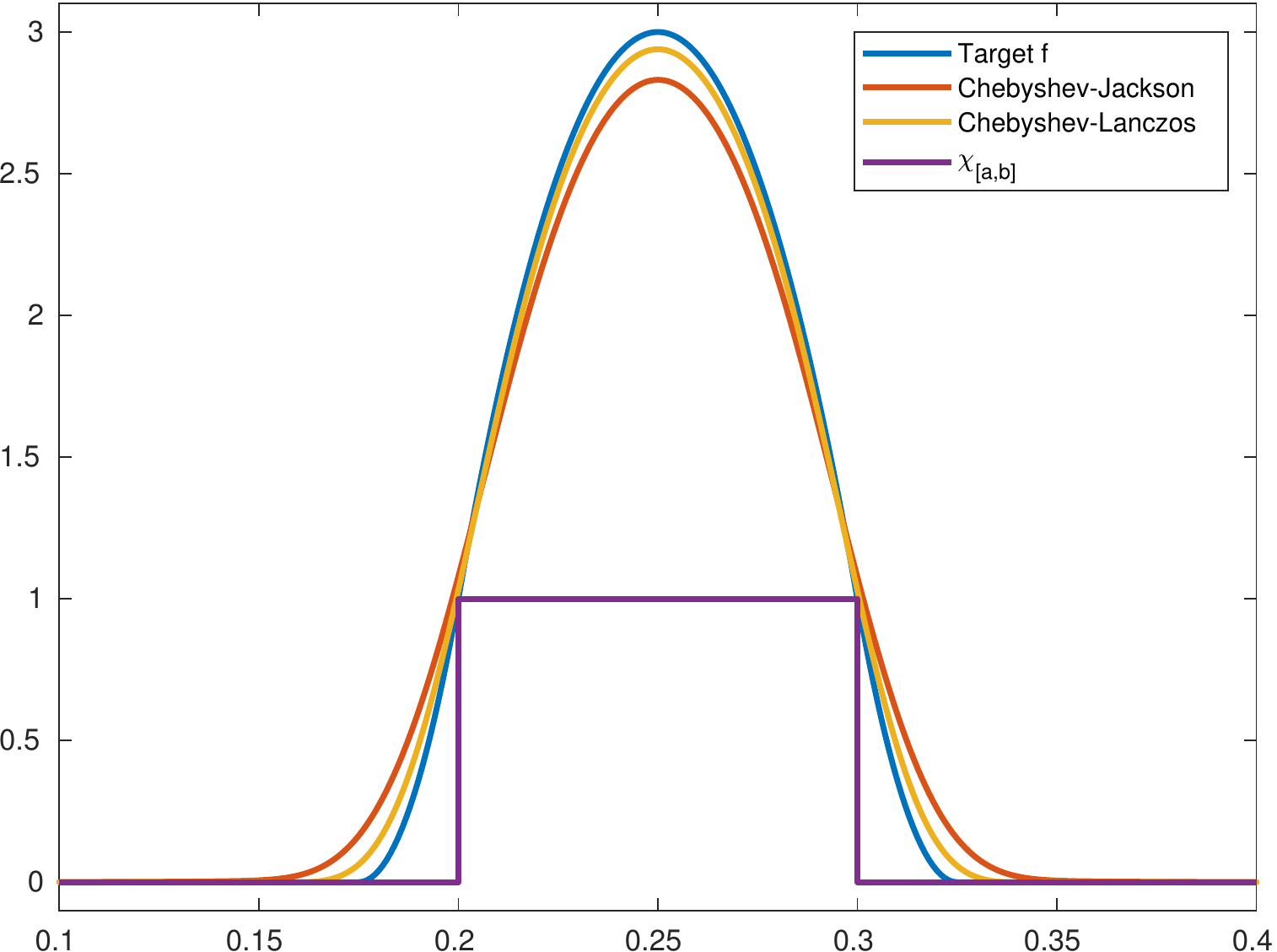}
  \caption{Plot of $p$ and Chebyshev-Jackson and Chebyshev-Lanczos
    polynomial expansions of degree~$k=200$ for $\xi=0.2$, $\eta=0.3$ and
    $\varepsilon=0.025$.  The violet curve shows the characteristic function
    $\chi_{[\xi,\eta]}$.  The values of the Chebyshev-Jackson and
    Chebyshev-Lanczos polynomials at the critical points are, respectively,
    $\pi_k^{J}(\xi-\varepsilon) \approx 0.164$,
    $\pi_k^{J}(\eta+\varepsilon) \approx 0.154$,
    $\pi_k^{L}(\xi-\varepsilon) \approx 0.0624$,
    $\pi_k^{L}(\eta+\varepsilon) \approx 0.0578$.}
  \label{fig:three_plots}
\end{figure}
The Chebyshev--Jackson (Chebyshev polynomial expansion with Jackson
smoothing) $\pi_k^{J}(x)$, for $x=\xi-\varepsilon$ and $\eta+\varepsilon$,
converges monotonically to zero
($=p(\xi-\varepsilon) = p(\eta+\varepsilon)$) from above.  We can therefore
determine the polynomial degree $k$ such that
\begin{equation}  \label{eq:thresh}
  \max\{\pi_k^{J}(\xi-\varepsilon), \pi_k^{J}(\eta+\varepsilon)\}
  \le r \min\{\pi_k^{J}(\xi), \pi_k^{J}(\eta)\}.
\end{equation}
A typical value for the \emph{convergence rate} $r$ is $0.1$.  We compare
with the values of $\pi_k$ at $\xi$ and $\eta$ as they can be away from~1 for
low degrees~$k$.  We check at both ends of the interval as the Chebyshev
polynomials are not symmetric.
If~\eqref{eq:thresh} is satisfied then the components in the directions of
eigenvalues \emph{outside} $(\xi-\varepsilon, \eta+\varepsilon)$ will decay
with a rate below $r$~\cite{parl:80, saad:11}.  So, the polynomial filter
takes care of these directions, while the Rayleigh--Ritz procedure handles
the directions associated with eigenvalues in
$(\xi-\varepsilon,\xi) \cup (\eta,\eta+\varepsilon)$.  For this to work, the
search space in the subspace iteration must be at least as large as the
number of eigenvalues in $(\xi-\varepsilon, \eta+\varepsilon)$.

By means of $\varepsilon$ and $r$ we can control the dimensions of subspaces
and the number of iteration steps.  The order of the filter polynomials
increases if $\varepsilon$ and $r$ decrease.  The order of the filter
polynomial decreases if we switch from Jackson to Lanczos smoothing.  If the
subspace dimension gets to large we may split $(\xi,\eta)$ in smaller
pieces.  This so-called `spectrum slicing' entails higher polynomial
degrees~\cite{lxvy:16}.



\section{Implementation}
\label{sec:impl}

We combined the methods discussed in the previous section and a few other
useful tools into a utility that can be employed to compute the eigenpairs
of a $n\times n$ real symmetric (or complex Hermitian) matrix $\bm{H}$
within a specified interval of interest $[\xi,\eta]$ in parallel by simply
providing an XML configuration file~\cite{acca:17}.  The outline of the
utility is displayed in Algorithm~\ref{alg:bp_utility}.
\begin{algorithm}[htb]
  \caption{The BosonPeak Utility}
  \label{alg:bp_utility}
  \begin{algorithmic}[1]
    \STATE Import user-specified configuration via XML file.

    \STATE Import matrix $\bm{H}$.

    \STATE \textbf{if} \emph{requested} \textbf{then} estimate extremal
    eigenvalues $\lambda_{\min}, \lambda_{\max}$ of $\bm{H}$ using a few
    Lanczos steps.

    \STATE Transform the matrix $\bm{H}$ such that
    $-1 \le\lambda_{\min} < \lambda_{\max} \le 1$, according
    to~\eqref{eq:transf.1}.
 
    \STATE \textbf{if} \emph{requested} \textbf{then} estimate the number of
    eigenvalues in the specified interval $[\xi,\eta]$.

    \STATE Compute the polynomial filter $\rho_k$ or $\pi_k$.
 
    \STATE Compute the eigenpairs $({\lambda}_j, \bm{u}_j)$,
    $\norm{\bm{u}_j}=1$, of $\bm{H}$ with
    ${\lambda}_j \in [\xi, \eta]$ and residual norms
    $r_j = \|\bm{H}\bm{u}_j - {\lambda}_j\bm{u}_j\| <\epsilon$ using
    either the thick restart Lanczos algorithm or subspace iteration.
    \label{alg:bp_utility_line_eigensolver}
    \STATE \textbf{if} \emph{requested} \textbf{then} save eigenvalues
    ${\lambda}_j$, associated eigenvectors $\bm{u}_j$, and residual norms
    $r_j$ to disk.
  \end{algorithmic}
\end{algorithm}
The utility is written in C++11 and uses Trilinos~\cite{Trilinos-Web-Site}
extensively.  Trilinos%
\footnote{\url{https://trilinos.org/packages/}} is a collection of
open-source software libraries, called \emph{packages}, for the
development of scientific applications.

The most basic Trilinos package is Epetra that provides classes for the
construction and use of sequential and distributed parallel linear algebra
objects.  The Trilinos solver packages are designed to work with Epetra
objects.  The most used linear algebra objects in our implementation are (i)
sparse matrices stored as \texttt{Epetra\_CrsMatrix} objects in the
compressed row storage (CRS) scheme, and (ii) \texttt{Epetra\_MultiVector}
objects that represent \emph{multivectors}, i.e., collections of dense
vectors.  Each vector in an \texttt{Epetra\_MultiVector} object is stored as
a contiguous array of double-precision numbers.  Both objects are
extensively used for sparse matrix-vector multiplications in the various
Trilinos solver packages.  All Trilinos packages resort to a method called
\texttt{Epetra\_Operator::Apply} to multiply a matrix with a (multi)vector.
Our operators are mostly matrix polynomials, and a call to
\texttt{Epetra\_Operator::Apply} entails the invocation of
Algorithm~\ref{alg:chebyshev_series}.

\emph{Anasazi}~\cite{bhlt:09} is a package that offers a collection of
algorithms for solving large-scale eigenvalue problems.  As part of the
package it provides solver managers to implement strategies for that
purpose.
We employ Anasazi's block Krylov--Schur eigensolver with thick restarts.
The subspace iteration that we discuss below is not a part of Anasazi.  We
implemented it ourselves, based on \texttt{Epetra} data structures.

The \emph{Teuchos} package is a collection of common tools used throughout
Trilinos.  Among other things, it provides templated access to BLAS and
LAPACK interfaces, parameter lists that allow to specify parameters for
different packages, and memory management tools for aiding in correct
allocation and deletion of memory.

Part of Teuchos' memory management tools is an implementation of a smart
Reference-Counted Pointer (RCP) class, which for an object tracks a count of
the number of references to it held by other objects.  Once the counter
becomes zero, the object can be deleted.  The advantage of a {RCP} is that
the possibility of memory leaks in a program can be reduced.  This is
important when working with rather large objects, e.g., an
\texttt{Epetra\_CrsMatrix} object storing over $10^9$ nonzero entries.
{RCP} objects are heavily used throughout our implementation to manage large
objects, especially large temporary objects that are only needed during a
fraction of the whole computation.

Trilinos supports distributed-memory parallel computations through the
Message Passing Interface (MPI).  Both the \texttt{Epetra\_CrsMatrix} and
the \texttt{Epetra\_Multivector} objects can be used in a distributed memory
environment by defining data distribution patterns using
\texttt{Epetra\_Map} objects.

The entries of a distributed object (such as rows or columns of an
\texttt{Epetra\_CrsMatrix} or the rows of an \texttt{Epetra\_Multivector})
are represented by \emph{global indices} uniquely over the entire object.  A
map essentially assigns global indices to available \textsc{MPI} ranks,
which in our case corresponds to a single core of a processor.


For the addressing, local and global indices in Epetra use by default the
$32$-bit \texttt{int} type.  Since our implementation is based on the C++11
language standard and we want to allow computations with large matrices, we
explicitly use $64$-bit global indices of type \texttt{long long} when
working with distributed linear algebra objects.  (Local indices are of type
\texttt{int}.)

An \texttt{Epetra\_Map} object encapsulates the details of distributing data
over \textsc{MPI} ranks.  \sloppy{In our implementation, we use \emph{contiguous}
and \emph{one-to-one} maps for the block row-wise distribution of the 
\texttt{Epetra\_CrsMatrix} and \texttt{Epetra\_MultiVector}
objects. \emph{Contiguous} means that the list of global indices on each
\textsc{MPI} rank forms an interval and is strictly increasing.}  A
\emph{one-to-one} map  allows a global index  to be owned by  a
single rank.  For the columns, the distribution pattern we are using
distributes the complete set of global column indices for a given global
row, meaning that if a rank $p$ owns the global row index $i$, it also owns
all global column indices $j$ on that row, thus having local access to the
global entry $(i,j)$.  The map used for the distribution of the columns is
thus not a one-to-one map, since a global column index can be owned by
multiple ranks.

The \emph{matrix import} implemented in the utility allows to efficiently
import large matrices stored in a \textsc{HDF5}%
\footnote{{\url{https://portal.hdfgroup.org/}}} file directly to an
\texttt{Epetra\_CrsMatrix} object.
\textsc{HDF5} is a data model, library, and file format for storing and
managing data collections of all sizes and complexity.  One of the
advantages of using the HDF5 file format to store and import large matrices
is the possibility to use MPI to read the HDF5 files in parallel.  For this
reason Trilinos provides the \texttt{EpetraExt::HDF5} class for importing a
matrix stored in a HDF5 file to a \texttt{Epetra\_CrsMatrix}.

Since the \texttt{EpetraExt::HDF5} class currently does not provide an
import function for matrices with $64$-bit global indices of type
\texttt{long long}, we extended the class by this functionality.  The
BosonPeak utility also provides a Python script that can be used to convert
matrices stored in the MatrixMarket format%
\footnote{\url{https://math.nist.gov/MatrixMarket/}} to a \textsc{HDF5} file
suitable for import.  The utility is described in detail in~\cite{acca:17}.

We now discuss two ways of computing the eigenvectors associated with the
eigenvalues of a prescribed interval $[\xi, \eta]$.
The first is an implementation of the thick-restart Lanczos algorithm
applied to $\rho_k(\bm{H})$.  We follow closely the implementation in EVSL,
see~\cite{lxvy:16}.  $\rho_k(t)$ is an approximation of the Dirac delta
distribution $\delta(t-\gamma)$ in $\bbbp_k$ as given
in~\eqref{eq:cheby-delta.1}.  $\gamma \approx (\xi+\eta)/2$ is chosen such
that $\rho_k(\xi) = \rho_k(\eta) = \tau$.  
%
The smaller $\tau$ (relative to $\rho_k(\gamma)$) the more prominent are the
eigenvalues $\rho_k(\lambda_j)$ for $\lambda_j \in [\xi, \eta]$ compared
with the eigenvalues outside of this interval.  Thus, the Lanczos algorithm
will extract these eigenvalues quickly.
The Lanczos algorithm also rapidly finds eigenvalues at the other end of the
spectrum, $\rho_k(\lambda_j)\approx0$.  But those are easily identified and
discarded.
Eigenpairs $(\mu_j, \bm{v}_j)$ of $\rho_k(\bm{H})$ with $\mu_j\ge \gamma$
are related to some eigenpair $(\lambda_{j'}, \bm{u}_{j'})$ of $\bm{H}$.
Certainly $\mu_j = \rho_k(\lambda_{j'})$.  Unless $\mu_j$ is simple,
$\bm{v}_j$ may not be an eigenvector of $\bm{H}$.  The reason for this is
that it can happen that $\mu_j = \rho_k(\lambda_{j'}) = \rho_k(\lambda_{j''})$,
$\lambda_{j'} \neq \lambda_{j''}$, in which case $\bm{v}_j$ may be a linear
combination of the eigenvectors  $\bm{u}_{j'}$ and $\bm{u}_{j''}$, and
possibly others.  The best way out of this problem is to execute a
Rayleigh--Ritz procedure~\cite{parl:80} that involves all eigenvectors
$\bm{v}_j$ of $\rho_k(\bm{H})$ associated with eigenvalues $\ge\tau$.
We employ the Lanczos algorithm with blocks of size 16, 32, or 64.  The
dimension of the search space is limited by $3n_{\text{ev}}$, where
$n_{\text{ev}}$ is the number of desired eigenvalues.  We consider
eigenpairs converged if
$\norm{\rho_k(\bm{H}) \bm{u}_i - \rho_k(\lambda_i)\bm{u}_i} < \epsilon
\norm{\rho_k(\bm{H})} \norm{\bm{u}_i} \le \epsilon
\rho_k(\gamma)\norm{\bm{u}_i}$
with $\epsilon=10^{-6}$.  This does not imply that
$\norm{\bm{H} \bm{u}_i - \lambda_i\bm{u}_i} < \epsilon \norm{\bm{H}}
\norm{\bm{u}_i}$,
though.  

The second approach to compute the eigenvectors associated with eigenvalues
in $[\xi, \eta]$ is based on a subspace iteration with the matrix
$\pi_k(\bm{H})$.  Note that in Algorithm~\ref{alg:sirra} $\bm{H}$ should be
substituted by $\pi_k(\bm{H})$.
\begin{algorithm}[htb]
  \caption{Subspace Iteration complemented by the Rayleigh--Ritz procedure}

  \label{alg:sirra}

  \begin{algorithmic}[1]
    \REQUIRE Symmetric positive definite matrix $\bm{H}\in\bbbr^{n\times n}$;
    number $p$ of desired eigenpairs; initial subspace
    $\bm{X}_0\in\bbbr^{n\times q}$ with $q\ge p$; tolerance $\tau$.

    \ENSURE $n\times p$ matrix $\bm{X}$ and $\bm{\Lambda} = \diag\{\lambda_1,
    \ldots, \lambda_p\}$ with $\lambda_1 \ge \cdots \ge \lambda_p$ such that 
    \begin{displaymath}
      \norm{\bm{H} \bm{x}_i - \lambda_i\bm{x}_i} < \tau,
      \qquad 1 \le i \le p.
    \end{displaymath}


    \STATE Orthonormalize starting vectors  $\bm{X}_0$.%
    \COMMENT{Gram--Schmidt orthogonalization by QR factorization}

    \FOR{$k = 1,\ldots$}

    \IF{$k\equiv1\Mod{n_{\text{RR}}}\ \&\ k>1$}

    \STATE   $\bm{Y}_{k}=\bm{HX}_{k-1}$;
    
    \STATE   $\bm{R}=\bm{Y}_{k} - \bm{X}_{k-1}\bm{\Lambda}_{k-1}$;%
    \COMMENT{Residual $\bm{R} = [\bm{r}_1,\ldots,\bm{r}_q]$}
    
    \STATE  $n_{\text{conv}} = \abs{\{1\le i\le q\ |\ \norm{\bm{r}_i}<\tau \}}$
    \COMMENT{Number of converged eigenpairs}

    \STATE \textbf{if} $n_{\text{conv}}\ge p$ \textbf{then} return\,($\bm{X} =
    \bm{X}_k$, $\bm{\Lambda} = \bm{\Lambda}_k$); 
    \textbf{end if}

    \STATE   $\bm{X}_{k} = \bm{Y}_{k}$;

    \ELSIF{$k\equiv-1\Mod{n_{\text{RR}}}$}
    \STATE   $\bm{X}_{k}\bm{R}_k = \bm{HX}_{k-1}$;
    \COMMENT{Iteration step with basis orthogonalization (QR)}

    \ELSIF {$k\equiv0\Mod{n_{\text{RR}}}$}
    \STATE $\bm{Y}_k = \bm{HX}_{k-1}$;
    \STATE $\bm{X}_k\bm{R}_k = \bm{Y}_k$;\COMMENT{Orthonormalize $\bm{Y}_k$ (QR
      factorization)}

    \STATE $\bm{U}_k \bm{\Lambda}_{k} \bm{V}_{k}^* = \bm{R}_k$;%
    \COMMENT{Singular value decomposition}

    \STATE $\bm{X}_k = \bm{X}_k \bm{U}_{k}$;
    \COMMENT{Ritz vectors}

    \ELSE
    \STATE $\bm{X}_{k} = \bm{HX}_{k-1}$;
    \COMMENT{Simple iteration step}

    \ENDIF
    \ENDFOR

  \end{algorithmic}
\end{algorithm}
The degree $k$ of $\pi_k$ should be the smallest integer for which
condition~\eqref{eq:thresh} is satisfied.  If $r=0.1$ and the search
subspace is not smaller than $q = \mu_{[\xi-\varepsilon,\eta+\varepsilon]}$,
the number of eigenvalues of $\bm{H}$ in
$[\xi-\varepsilon,\eta+\varepsilon]$, then convergence is rapid: the angle
between Ritz vectors and corresponding eigenvectors decreases per iteration
step by the factor~$r$.  Also the residual norm, which defines our stopping
criterion, decreases rapidly.  One could try to determine the number of
iteration steps in advance, based on $r$ and $\varepsilon$.
Note that $r$ can be arbitrarily close to~1 if $\varepsilon\approx0$.  The
larger $\varepsilon$, the smaller $k$ and the bigger
$\mu_{[\xi-\varepsilon,\eta+\varepsilon]}$.  The former is proportional to
work; the latter determines the size of the search space.  In subspace
iteration the memory space for two multivectors of size
$n\times \mu_{[\xi-\varepsilon,\eta+\varepsilon]}$ are needed.  More memory
is required in Algorithm~\ref{alg:chebyshev_series}.  The actual amount
depends on the block size.

Algorithm~\ref{alg:sirra} follows quite closely the one proposed by
Rutishauser~\cite{ruti:69}, see also~\cite[Table\,14.4]{parl:80}.  The positive integer
$n_{\text{RR}}$ indicates the frequency with which the Rayleigh--Ritz
procedure is applied.
In the iteration step before the RR procedure, the basis vectors in
$\bm{X}_k$ are orthonormalized.  In the iteration step after the RR
procedure, the convergence criterion is applied.
%
In our experiments we set $n_{\text{RR}}=6$ which, combined with
$\tau=10^{-6}$, means that we have convergence after $n_{\text{RR}}$
iteration steps.

Considerable savings can be achieved by \emph{locking} converged Ritz
vectors.  If $n_{\text{conv}}< p$ vectors have converged then only the
columns $n_{\text{conv}}\!+\!1, \ldots, q$ have to be iterated.  So, in
lines~10, 12, and~17, a reduced number of columns have to be multiplied by
$\bm{H}$.  In order to profit from locking we increase $r$, setting
$r=\sqrt{.1}$ or $r=\sqrt[4]{.1}$.  This means that we have convergence only
after $2n_{\text{RR}}$ or $4n_{\text{RR}}$ iteration steps.  However, the
degree of the polynomial smoother is reduced considerably such that,
combined with locking, shorter execution times can result.
\section{Numerical experiments}
\label{sec:experiments}

\subsection{Comparisons by means of synthetic problems}

To compare solvers and filters we first limit ourselves to synthetic
diagonal eigenvalue problems.  We construct diagonal matrices $\bm{H}$ that
have a prescribed eigenvalue distribution.  The eigenvalues in intervals
$(\xi, \eta)$ are then computed by either subspace iteration or
thick-restart Lanczos, both with polynomial filtering.  We know in advance
how many eigenvalues we are looking for, whence we do not estimate this
number.  The filter maps these $p$~eigenvalues to the largest of the
filtered matrix.  In subspace iteration (SI) we iterate with a block
size~$q$ which is initially the number of eigenvalues of $\bm{H}$ in
$(\xi-\varepsilon,\eta+\varepsilon)$.  $q$~decreases as the number of found
eigenpairs increases.  In the block Lanczos algorithm the block size is
fixed to~8.  The block Lanczos algorithm with thick restart (TRLanczos) is
implemented in Anasazi as a variant of the block Krylov--Schur (BKS)
algorithm~\cite{bhlt:09}.
Note that we use the notions TRLanczos and BKS interchangeably.

We measure the computational effort by matrix-vector multiplications
(MatVec's).  In realistic computations the MatVec's dominate Rayleigh--Ritz
procedure or thick restart.  In both SI and BKS the computational effort is
the number of iteration steps times the degree of the polynomial filter
times the block size.


\begin{table}[htb]
  \centering
  \begin{tabular}{|c|c|c|c|c|c|c|}
    \hline
    &\multicolumn{6}{|c|}{SI + Piecewise quadratic polynomial filter} \\ \hline
    &\multicolumn{2}{|c|}{Jackson smoothing} &
    \multicolumn{2}{|c|}{Lanczos damping} &
    \multicolumn{2}{|c|}{No smoothing} \\ \hline
    & $d$ & \#MatVec's &$d$ & \#MatVec's & $d$ & \#MatVec's \\ \hline
    $r=0.1$ & 5'753 & 3'020'325 & 3'508 & 1'841'700 & 1'163 & 1'099'035 \\
    $r=\sqrt{0.1}$ & 2'739 & 2'670'525 & 1'771 & 1'726'725 & 893 & 870'675 \\
    $r=\sqrt[4]{0.1}$ & 1'548 & 2'616'120 & 1'058 & 1'788'020 & 673 & 1'137'370\\
    \hline\hline
    &\multicolumn{6}{|c|}{SI + Dirac polynomial filter} \\ \hline
    &\multicolumn{2}{|c|}{Jackson smoothing} &
    \multicolumn{2}{|c|}{Lanczos damping} &
    \multicolumn{2}{|c|}{No smoothing} \\\hline
    & $d$ & \#MatVec's &$d$ & \#MatVec's & $d$ & \#MatVec's \\ \hline
    $\phi=0.8$ & 942 & 3'706'770 & 696 & 2'724'840 & 494 & 1'822'860\\
    $\phi=0.6$ & 1'412 & 2'760'460 & 1'034 & 1'747'460 & 727 & 1'206'820\\
    \hline\hline
    &\multicolumn{6}{|c|}{BKS + Piecewise quadratic polynomial filter} \\ \hline
    &\multicolumn{2}{|c|}{Jackson smoothing} &
    \multicolumn{2}{|c|}{Lanczos damping} &
    \multicolumn{2}{|c|}{No smoothing} \\\hline
    & $d$ & \#MatVec's &$d$ & \#MatVec's & $d$ & \#MatVec's \\ \hline
    $r=0.1$ & 5'753 & 828'432 & 3'508 & 505'152 & 1'163 & {167'472} \\
    $r=\sqrt{0.1}$ & 2'739 & 394'416 & 1'771 & 255'024 & 893 & 207'176  \\
    $r=\sqrt[4]{0.1}$ & 1'548 & 359'136 & 1'058 & 245'456 & 673 & \textbf{156'136}  \\
    \hline\hline
    &\multicolumn{6}{|c|}{BKS + Dirac polynomial filter} \\ \hline
    &\multicolumn{2}{|c|}{Jackson smoothing} &
    \multicolumn{2}{|c|}{Lanczos damping} &
    \multicolumn{2}{|c|}{No smoothing} \\\hline
    & $d$ & \#MatVec's &$d$ & \#MatVec's & $d$ & \#MatVec's \\ \hline
    $\phi=0.8$ & 942 & 301'440 & 696 & 222'720 & 494 & 201'552 \\
    $\phi=0.6$ & 1'412 & 327'584 & 1'034 & 239'888 & 727 & {168'664} \\
    \hline\hline
  \end{tabular}
  \caption{Summary of test runs with subspace iteration and various
    polynomial filters.  Equally distributed eigenvalues.  $\xi=-0.9$,
    $\eta=-0.898$, $\varepsilon=0.0005$, $p=50$, $q=75$ in SI, BKS block
    size is 8.  Best performance is in bold.}
  \label{tab:diag.solve.1}
\end{table}
We consider three test cases, all with diagonal matrices of order
50'000.  In the first case the matrix $\bm{A}$ has equally distributed
eigenvalues in $[-1,1]$.  We computed the 50~eigenvalues in the interval
$(\xi, \eta)=(-0.9, -0.898)$.  In this test case as in later ones we start
the iterations with (the same) random multivector of the appropriate size.
The progress of the algorithm depends of course on the initial data.
However, the results given here represent typical behavior.  We did not
execute an exhaustive search for best parameters, though.
The results for the first test matrix with equidistant eigenvalues are
presented in Table~\ref{tab:diag.solve.1}.  The block Krylov--Schur
algorithm clearly outperforms subspace iteration, by a factor~5 when
  comparing the best parameter sets.  There is no clear winner among the
polynomial filters.  Both types of filters perform best without smoothing.
The reduction of the polynomial degree is larger than the gain by smoothing
or damping the Gibbs oscillations.  This holds particularly for the piecewise
quadratic polynomial filter.
The Dirac polynomial filter performs better with $\phi=0.6$ than with
$\phi=0.8$, i.e., with the better approximation of the Dirac distribution.
This leads to higher polynomial degrees, but to lower iteration counts.  On
the other hand, the piecewise quadratic polynomial filter prefers a weak
approximation of the step function, $r>0.1$, but the results are not
consistent.  With SI this increases the iteration count but decreases the
cost of a single iteration step and increases the potential for locking.

\begin{table}[htb]
  \centering
  \begin{tabular}{|c|c|c|c|c|c|c|}
    \hline\hline
    &\multicolumn{6}{|c|}{\rule{0pt}{10pt}SI + Piecewise quadratic polynomial filter} \\ \hline
    &\multicolumn{2}{|c|}{Jackson smoothing} &
    \multicolumn{2}{|c|}{Lanczos damping} &
    \multicolumn{2}{|c|}{No smoothing} \\ \hline
    & $d$ & \#MatVec's &$d$ & \#MatVec's & $d$ & \#MatVec's \\ \hline
    $r=0.1$ & 5'764 & 645'568 & 3'498 & 391'776 & 1'220 & 223'260 \\
    $r=\sqrt{0.1}$ & 2'663 & 527'274 & 1'702 & 336'996 & 662 & 134'386 \\
    $r=\sqrt[4]{0.1}$ & 1'389 & 486'150 & 904 & 316'400 & 485 & 169'750\\
    \hline\hline
    &\multicolumn{6}{|c|}{\rule{0pt}{10pt}SI + Dirac polynomial filter} \\
    \hline
    &\multicolumn{2}{|c|}{Jackson smoothing} &
    \multicolumn{2}{|c|}{Lanczos damping} &
    \multicolumn{2}{|c|}{No smoothing} \\\hline
    & $d$ & \#MatVec's &$d$ & \#MatVec's & $d$ & \#MatVec's \\ \hline
    $\phi=0.8$ & 472 & 722'160 & 349 & 512'681 & 249 & 360'801 \\
    $\phi=0.6$ & 708 & 477'192 & 519 & 347'211 & 365 & 221'920 \\
    \hline\hline
    &\multicolumn{6}{|c|}{\rule{0pt}{10pt}BKS + Piecewise quadratic polynomial filter} \\
    \hline
    &\multicolumn{2}{|c|}{Jackson smoothing} &
    \multicolumn{2}{|c|}{Lanczos damping} &
    \multicolumn{2}{|c|}{No smoothing} \\\hline
    & $d$ & \#MatVec's &$d$ & \#MatVec's & $d$ & \#MatVec's \\ \hline
    $r=0.1$ & 5'764 & 276'672 & 3'498 & 167'904 & 1'220 & 107'360 \\
    $r=\sqrt{0.1}$ & 2'663 & 149'128 & 1'702 & 95'312 & 662 & 74'144 \\
    $r=\sqrt[4]{0.1}$&1'389 & 88'896 & 904 & 65'088 & 485 & 62'080 \\
    \hline\hline
    &\multicolumn{6}{|c|}{\rule{0pt}{10pt}BKS + Dirac polynomial filter} \\
    \hline
    &\multicolumn{2}{|c|}{Jackson smoothing} &
    \multicolumn{2}{|c|}{Lanczos damping} &
    \multicolumn{2}{|c|}{No smoothing} \\\hline
    & $d$ & \#MatVec's &$d$ & \#MatVec's & $d$ & \#MatVec's \\ \hline
    $\phi=0.8$ & 472 & \textbf{30'208} & 349 & 41'880 & 249 & 47'808 \\
    $\phi=0.6$ & 708 & 45'312 & 519 & 53'976 & 365 & 70'080 \\
    \hline\hline
  \end{tabular}
  \caption{Summary of test runs with subspace iteration and various
    polynomial filters. 
    Eigenvalue density increasing towards $+1$.  $\xi=-0.9$, $\eta=-0.896$,
    $\varepsilon=0.0005$, 
    $p=10$, $q=16$ in SI, BKS block size is 8.  Best performance is in bold.}
  \label{tab:diag.solve.2}
\end{table}
The density of the eigenvalues of the matrix $\bm{H}$ of the second test case
increases from~$-1$ to~1.  In the interval that we chose,
$(\xi, \eta)=(-0.9, -0.896)$, there are only 10~eigenvalues.  These are
better separated from the rest of the spectrum than in the previous test
case.  The interval in this example is twice as wide as in the previous
example.  Therefore, the polynomial degrees are lower.  Note that the filter
polynomials only depend on the interval, not on the number of eigenvalues it
contains.

The results of the second test are given in Table~\ref{tab:diag.solve.2}.
Here, the thick restart Lanczos (BKS) algorithm with the Dirac filter
($\phi=0.8$) and Jackson smoothing is twice as fast as BKS with the
piecewise quadratic polynomial filter ($r=\sqrt[4]{0.1}$) without smoothing
and more than four times faster than any combination of subspace iteration.
%

\begin{table}[htb]
  \centering
  \begin{tabular}{|c|c|c|c|c|c|c|}
    \hline\hline
    &\multicolumn{6}{|c|}{SI + Piecewise quadratic polynomial filter} \\ \hline
    &\multicolumn{2}{|c|}{Jackson smoothing} &
    \multicolumn{2}{|c|}{Lanczos damping} &
    \multicolumn{2}{|c|}{No smoothing} \\ \hline
    & $d$ & \#MatVec's &$d$ & \#MatVec's & $d$ & \#MatVec's \\ \hline
    $r=0.1$ & 1'521 & 958'230 & 924 & 582'120 & 328  & 290'280 \\
    $r=\sqrt{0.1}$ & 709 & 825'985 & 454 & 528'910 & 180 & 209'700 \\
    $r=\sqrt[4]{0.1}$ & 373 & 630'370 & 244 & 412'360 & 130 & 230'480 \\
    \hline\hline
    &\multicolumn{6}{|c|}{SI + Dirac polynomial filter} \\ \hline
    &\multicolumn{2}{|c|}{Jackson smoothing} &
    \multicolumn{2}{|c|}{Lanczos damping} &
    \multicolumn{2}{|c|}{No smoothing} \\\hline
    & $d$ & \#MatVec's &$d$ & \#MatVec's & $d$ & \#MatVec's \\ \hline
    $\phi=0.8$ & 128 & 945'280 & 95 & 679'725 & 68 & 458'660 \\
    $\phi=0.6$ & 192 & 668'745 & 141 & 487'155 & 100 & 321'000 \\
    \hline\hline
    &\multicolumn{6}{|c|}{\rule{0pt}{10pt}BKS + Piecewise quadratic polynomial filter} \\
    \hline
    &\multicolumn{2}{|c|}{Jackson smoothing} &
    \multicolumn{2}{|c|}{Lanczos damping} &
    \multicolumn{2}{|c|}{No smoothing} \\\hline
    & $d$ & \#MatVec's &$d$ & \#MatVec's & $d$ & \#MatVec's \\ \hline
    $r=0.1$ & 1'521 & 292'032 & 924 & 177'408 & 328 & 102'336 \\
    $r=\sqrt{0.1}$ & 709 & 136'128 & 454 & 250'608 & 180 & 56'160 \\
    $r=\sqrt[4]{0.1}$ & 373 & 71'616 & 244 & 46'848 & 130 & 56'160 \\
    \hline\hline
    &\multicolumn{6}{|c|}{BKS + Dirac polynomial filter} \\ \hline
    &\multicolumn{2}{|c|}{Jackson smoothing} &
    \multicolumn{2}{|c|}{Lanczos damping} &
    \multicolumn{2}{|c|}{No smoothing} \\\hline
    & $d$ & \#MatVec's &$d$ & \#MatVec's & $d$ & \#MatVec's \\ \hline
    $\phi=0.8$ & 128 & 39'936 & 95 & \textbf{29'640} & 68 & \textbf{29'376} \\
    $\phi=0.6$ & 192 & 59'904 & 141 & 43'992 & 100 & 67'200 \\
    \hline\hline
  \end{tabular}
  \caption{Summary of test runs with subspace iteration and various
    polynomial filters. 
    Eigenvalue density increasing towards $+1$.  $\xi=-0.9$, $b=-0.885$,
    $\varepsilon=0.002$, $p=66$, $q=90$, BKS block size 8.  Best
    performance is in bold.}
  \label{tab:diag.solve.3}
\end{table}
In the third test case we chose the same matrix as in the second but we
expanded the search interval to have about the same number of eigenvalues as
in the first test case.  The interval $(\xi, \eta)=(-0.9, -0.885)$ encloses
66~eigenvalues that form a superset of the eigenvalues of the matrix in test
case~2.
While the eigenvalue count is more than 6~times higher, the degrees of the
filter polynomials decreases by a factor of almost~4.  This has the effect
that the costs with respect to test example~2 increase slightly for SI and
even decrease for BKS, see~Table~\ref{tab:diag.solve.3}.
The outcome is similar to case~2.  BKS with Dirac polynomial filtering
($\phi=0.8$) is faster than BKS with piecewise quadratic polynomial
filtering by a factor almost two and faster than any varian of SI by a
factor almost~7.  In contrast to test case~2, here no smoothing or Lanczos
damping perform better than Jackson smoothing.




Based on these results we decided to employ the Trilinos block Krylov--Schur
eigensolver combined with the Dirac polynomial filter with $\phi=0.8$ in the
real world problems treated below.  
We chose to employ Jackson smoothing as our real world problems most
resemble test case~2: we compute a few eigenvalues w.r.t.\ the problem size
and the eigenvalue density increases towards higher eigenvalues.

\subsection{A realistic application: a glassy structure}

For a preliminary usage of the developed algorithm, we explore the spectrum
of a set of instances of a glassy structure involving $N = 256'000$ atoms.
The atomic positions of this structure are produced through a series of
molecular dynamics simulations involving: a well-equilibrated liquid at
temperatures well above the melting temperature, a quench to the lower
temperatures of the amorphous solid regime, and a final relaxation which
brings the system to a local potential energy minimum from which the
dynamical matrices of order $3\! \times\! 256'000 = 768'000$ can be
calculated. 
Note that different equivalent initial distribution of the atoms lead to
different \emph{realizations} of the configuration.
The empirical atomic
interaction model used to perform these simulations is based on a
Lennard--Jones force model~\cite{wahn:91}, which describes the interaction
between atoms of two types differing in both size and mass.  Periodic
boundary conditions are used to remove the explicit structural effect of a
surface.  For the chosen density, the periodicity length is $L=101.714585$
where the unit distance is close to the average atomic bond length.  Details
of the sample preparation procedure and the resulting glassy structures may
be found in refs.~\cite{dema:17,dema:18,dema:18a}.  These
aspects entail a set of sparse dynamical matrices, $\bm{H}_i$, which have
$\sim\!250$ nonzero elements per row, cf.\ column~2 in Table~\ref{tab:1}.
The sparsity of the matrices is thus about $3\cdot10^{-4}$.  Each of these
matrices represents one realisation of a configuration.

Past work considering much smaller glassy atomic configurations using the
Lennard--Jones potential~\cite{deml:12} suggests that the eigenvalue regime
at which sound breaks up -- the so-called boson peak (PB) regime -- occurs
in the approximate $\lambda$-interval $[1,2]$, where $\lambda_{\max}$ was
around~1800.
Since the infinite-dimensional operator
underlying~\eqref{eq:intro.3} is bounded, we do not expect $\lambda_{\max}$
to increase much with increased system size.
%
Indeed, the matrices $\bm{H}_i$ all have their eigenvalues in the interval
$[0,1920]$.
We are looking for those in the two subintervals
$[\xi,\eta] = [0.1,1]$ and $[\xi,\eta] = [1,2]$ that are mapped by the
linear function~\eqref{eq:transf.1} to
$[-\frac{9599}{9600}, -\frac{9590}{9600}] \approx [-0.99990,
-0.99896]$
and
$[-\frac{9590}{9600}, -\frac{9580}{9600}] \approx [-0.99896,
-0.99792]$, respectively.
These computations were part of an exploration of the eigenstructure of
the $\bm{H}_i$ down to a value of $\lambda=0$.

\begin{table}[htb]
  \centering
  \begin{tabular}{|c|c||c|c|c|c||c|c|c|c||}
    \hline
    && \multicolumn{3}{c|}{ $(\xi,\eta) = (0.1,1)$} & time & \multicolumn{3}{c|}{ $(\xi,\eta) = (1,2)$} & time \\
    && \multicolumn{3}{c|}{ $\text{degree}(\rho_k) = 121$} & to & \multicolumn{3}{c|}{ $\text{degree}(\rho_k) = 156$} & to \\
    Matrix  & nnz($\bm{H}_i$) & \multicolumn{3}{c|}{\# EVs} & sol'n & \multicolumn{3}{c|}{\# EVs} & sol'n\\ 
    && true & est. & req &  [sec] & true & est. & req &  [sec] \\\hline
    $\bm{H}_1$ & $191'893'806$ & 249 & 259 & 300 & $1'225$ & 819 & 798 & 900 & $4'059$ \\ \hline
    $\bm{H}_2$ & $191'883'888$ & 236 & 233 & 300 & $1'224$ & 794 & 780 & 900 & $3'637$ \\ \hline
    $\bm{H}_3$ & $191'903'166$ & 239 & 249 & 300 & $1'243$ & 811 & 793 & 900 & $4'551$ \\ \hline
    $\bm{H}_4$ & $191'851'848$ & 249 & 262 & 300 & $1'242$ & 826 & 812 & 900 & $4'600$ \\ \hline
    $\bm{H}_5$ & $191'832'588$ & 249 & 259 & 300 & $1'196$ & 828 & 807 & 900 & $4'094$ \\ \hline
    $\bm{H}_6$ & $191'859'012$ & 259 & 262 & 300 & $1'218$ & 820 & 817 & 900 & $4'169$ \\ \hline
    $\bm{H}_7$ & $191'887'542$ & 238 & 248 & 300 & $1'220$ & 820 & 796 & 900 & $4'159$ \\ \hline
    $\bm{H}_8$ & $191'853'378$ & 245 & 259 & 300 & $1'206$ & 806 & 806 & 900 & $4'146$ \\ \hline
  \end{tabular}
  
  \caption{Results for the partial eigenvalue computations of the Hessians
    $\bm{H}_{1}, \ldots, \bm{H}_{8}$ with $N = 256'000$ atoms, i.e., order
    $n=768'000$. 
    $\text{nnz}(\bm{H}_i)$ denotes the number of nonzeros of $\bm{H}_i$. 
    %
    The time to
    solution shows the time of the eigensolver part of the utility (line
    \ref{alg:bp_utility_line_eigensolver} in Algorithm
    \ref{alg:bp_utility}).  It is with respect to the
    computations with 48 cores on Euler II.
  }
  \label{tab:1}
\end{table}

A survey of results for the partial eigenvalue computations of the
Hessians $\bm{H}_{1}, \ldots, \bm{H}_{8}$ 
is given in Table~\ref{tab:1}.
For each matrix $\bm{H}_{i}$ and each interval $[\xi, \eta]$ we give the
true and estimated numbers of eigenvalues in the respective interval and the
time to compute the former.  The true numbers of eigenvalues are obtained as
a result of the computations that targeted at 300 eigenvalues for the left
interval and 900 for the right interval.  These are $10-20\%$ more
eigenvalues than estimated and, as such, a crude upper bound for
$\mu_{[\xi, \eta]}$.
%
The estimates have been obtained by the technique discussed in
subsection~\ref{subsec:counting} with $M=30$ samples
in~\eqref{eq:eigval_count_estimator} and degree $k=100$
for which parameters de Napoli et al.~\cite{naps:16} report very good
results.  One has to be careful when choosing $k$, though.  $\psi_k$ has to
be a reasonably good approximation of $\chi_{[\xi,\eta]}$.  Otherwise the
estimation may be completely off.  Generally, $k$ should increase as the
width of $[\xi, \eta]$ shrinks.

We computed the eigenvalues of the $\bm{H}_i$ with Anasazi's block
Krylov--Schur (BKS) algorithm (in fact, the thick restarted block Lanczos
algorithm) with the same parameter values.  The block size was fixed at~4.
The number of blocks was limited to~225.  The maximal dimension of the
Krylov space was set to $3\times n_{\text{ev}}$, where $n_{\text{ev}}$
is the number of desired eigenvalues.  The number of blocks was thus limited
to $\frac{3 \times n_{\text{ev}}}{\text{block size}}$.
The threshold was set to $\tau=0.9$.

We have implemented the operator $\rho_k(\bm{H}_i)$ in Trilinos.  The degree
$k$ of the filter polynomial is given in Table~\ref{tab:1}.
One call of the operator amounts to the execution of $k$
matrix-multivector multiplications.  Remember that the block size is~4.
These $4k$ MatVec's constitute more than 99\% of the execution
time of the solver.

The computations were carried out on Euler~II of ETH Zurich's
compute cluster%
\footnote{\url{https://scicomp.ethz.ch/wiki/Euler\#Euler\_II}}.
Euler~II comprises 768~compute, each equipped with two 12-core Intel Xeon
E5-2680v3 processors (2.5-3.3\,GHz) and between 64 and 512\,GB of DDR4
memory.  Euler II also contains 4 very large memory nodes each equipped with
four 16-core Intel Xeon E7-8867v3 processors (2.5\,GHz) and 3072\,GB of DDR4
memory.

In the specified environment, our implementation worked with OpenMPI\,1.65,
HDF5\,1.8.12, Boost\,1.57.0, and Trilinos\,12.2.1.  Further, the code has been
compiled with GCC\,4.8.2 and the following optimization flags:
\begin{center}
  \texttt{-ftree-vectorize -march=corei7-avx -mavx -std=c++11 -O3}.
\end{center}
The convergence criterion requires that the residual norms
\begin{displaymath}
  \norm{\rho_k(\bm{H}_i)\bm{u}_j - \mu_j \bm{u}_j} <
  \epsilon,
  \qquad \epsilon=10^{-6},\quad  \norm{\bm{u}_j}=1,
\end{displaymath}
for all \emph{requested} eigenpairs $(\mu_j, \bm{u}_j)$.  This lead to
very accurate eigenpairs of the original matrices,
\begin{displaymath}
  {r}_j \equiv \norm{\bm{H}_i\bm{u}_j - \lambda_j \bm{u}_j} <
  3.5\cdot10^{-8},\quad  \norm{\bm{u}_j}=1,
\end{displaymath}
or better for all desired (true) eigenpairs $(\lambda_j, \bm{u}_j)$.  Since
we compute too many eigenpairs, the desired ones are finally too accurate.
So, it is important for good efficiency to have accurate estimates for the
eigenvalue counts.  A high security margin entails high computational and
memory costs, together with overly accurate results.
The times to find the 300 largest eigenvalues and associated eigenvectors of
$\rho_k(\bm{H}_i)$ in the interval $[0.1,1]$ are in the average $1'222$\,sec
with small variance.  The 900 eigenpairs in the interval $[1,2]$ take about
$4'177$\,sec to compute with larger variance.  We attribute the large
variance to the heterogeneous cluster and the larger execution times.

\subsection{Scalability study}

To test the parallel scalability of our code we consider a
much larger glassy structure comprising of $1'372'000$ atoms.
The Hessian matrix has now order $n=4'116'000$ and $1'028'329'164$ nonzero
entries.  The number of nonzeros per row is again about~250, leading to a
sparsity of $6\cdot10^{-5}$.

In Table~\ref{tab:large} we display the computational results for five
subintervals of the investigated interval
$[1,2] \subset [0,\lambda_{\max}(\bm{H})]$ with
$\lambda_{\max}(\bm{H}) = 1941$. 
These computations were carried out on
Euler~V, the newest extension of ETH Zurich's compute cluster%
\footnote{\url{https://scicomp.ethz.ch/wiki/Euler\#Euler\_V}}.
Euler~V contains 352 compute nodes (Hewlett-Packard BL460c Gen10), each
equipped with two 12-core Intel Xeon Gold 5118 processors (2.3 GHz nominal,
3.2 GHz peak) and 96 GB of DDR4 memory clocked at 2400 MHz.  We consider
Euler~V a homogeneous cluster.
\begin{table}[htb]
  \centering
  \begin{tabular}{|c|c|c|c|c|c|c|c|c|c|c|c|c|}
    \hline\hline
    blk&\#    &dim   &re- & \# blk & poly& \# &$\xi$&$\eta$ & evs&  evs&  evs & time\\
    size& blks&$\mathcal{K}_d$& starts & steps &degree  & MatVec's &    &       & est  & req & true&[sec]\\
    \hline\hline
    8 & 21 & 168 & 3 & 63 & $10'142$ & $5'111'568$ & 1.169 & 1.183 &  49 &  55 &  46 & $22'667$ \\
    8 & 21 & 168 & 3 & 77 & $10'202$ & $6'284'432$ & 1.183 & 1.197 &  49 &  55 &  48 & $27'959$ \\
    8 & 21 & 168 & 4 & 91 & $10'262$ & $7'470'736$ & 1.197 & 1.211 &  50 &  55 &  54 & $32'937$ \\
    8 & 48 & 384 & 2 & 80 & $ 5'175$ & $3'312'000$ & 1.211 & 1.239 & 101 & 128 & 106 & $14'647$ \\
    8 & 48 & 384 & 2 & 80 & $ 5'233$ & $3'349'120$ & 1.239 & 1.267 & 103 & 128 & 105 & $14'850$ \\
    \hline\hline
  \end{tabular}
  \caption{Computational results for large glassy structure.
    $\xi$, $\eta$ are given with reference to the interval $[0,1941]$.
    The number of matrix-vector multiplications (\#MatVec) equals (blk
    size)$\times$(\#blk steps)$\times$(poly degree).  Times are for
    360~cores on Euler~V.}
  \label{tab:large}
\end{table}

\begin{table}[htb]
  \centering
  \begin{tabular}{|c|c|c|c|c|c|c|c|}
    \hline\hline
    evs &  evs & deg & \multicolumn{2}{|c|}{$p=360$} & \multicolumn{2}{|c|}{$p=720$} & `speedup' \\
    conf& found && \#blk steps & time & \#blk steps & time & \\
    \hline\hline
	55 &  46  & $10'142$ & 63 & $22'667$ & 63 & $11'570$ & 1.96 \\
	55 &  48  & $10'202$ & 77 & $27'959$ & 63 & $11'506$ & 2.43 \\
	55 &  54  & $10'262$ & 91 & $32'937$ & 77 & $14'229$ & 2.31 \\
	128 & 106 & $ 5'175$ & 80 & $14'647$ &112 & $10'570$ & 1.39 \\
	128 & 105 & $ 5'233$ & 80 & $14'850$ &112 & $10'547$ & 1.41 \\
    \hline\hline
  \end{tabular}
  \caption{Execution times for $p=360$
    and $p=720$ cores for the intervals
    in Table~\ref{tab:large} and derived speedups.}
  \label{tab:speedups}
\end{table}

There are three intervals of length 0.014 with about 50 eigenvalues and
two intervals of length 0.028 with about 105 eigenvalues.  We chose equal
block size~8 for all computations.
The number of blocks was again limited to
$\frac{3 \times n_{\text{ev}}}{\text{block size}}$ with $n_{\text{ev}}$
the number of requested eigenvalues.
The maximal dimension of the Krylov space equals the block size times the
number of blocks.  If it is attained then a restart is issued.
The number of restarts is obtained by the number of block steps divided by
the number of blocks.

Interestingly, it is faster to compute the 105 eigenvalues in the longer
intervals than the 50 eigenvalues in the shorter ones, essentially because
of the lower degrees of the filter polynomials.  Also the maximal
dimension of the search space is relatively larger.  This eases the
extraction of the desired information.  The overhead due to
reorthogonalizations is negligible.  The times in Table~\ref{tab:large}
are very well in proportion with the number of MatVec's.  We consistently
observe 226\,MatVec's per sec.  The times given are the fastest out of
three runs.

In Table~\ref{tab:speedups} we replicate the execution times of
Table~\ref{tab:large} for 360~cores and complement them with the execution
times obtained with 720 cores.  360~cores correspond to 15~nodes; 720~cores
correspond to 30~nodes of Euler~V.  Evidently, the runs with 720 cores should
take only half the time of the ones on 360 cores, amounting to a speedup
of two.  Since most of the computing time is spent in (blocked)
matrix-vector multiplications a speedup close to two can indeed be
expected.
The data is distributed among the cores in the standard block
row-wise fashion of Trilinos.  The computations are similar, in
particular, the initial vectors are equal.  Nevertheless, the number of
iteration steps until convergence can differ significantly.  Therefore,
the speedup appears to be erratic.  However, if we compare the execution
times of the single block steps, then we observe speedups close to two,
more precesely between~1.94 and~1.988.
With 720 cores about 443\,MatVec's are executed per second.

\subsection{Physics results}

In what follows, the eigenmode $\bm{q}$ is seen as consisting of $N$
3-vectors, $\bm{q}_{i} = (q_{i}^{1}, q_{i}^{2}, q_{i}^{3})$, where the
superscript indicates the coordinate direction.   To gain an estimate of
the number of atoms involved in a normalized eigenmode the participation
ratio~\cite{bede:70} is used,
\begin{equation} \label{eq:PR_def}
  \mathrm{PR}=\frac{1}{\sum_{i}|\bm{q}_{i}|^{4}}
  = \frac{1}{\norm{\bm{q}}_4^4},
\end{equation}
where $\norm{\bm{q}}_2=1$ is assumed.  When an eigenmode has constant
values, say $|\bm{q}_{i}|=1/\sqrt{N}$ then $\mathrm{PR}=1$ and all atoms
are said to partake in the eigenmode.  On the other hand when
$\abs{\bm{q}_{i}}=1$ for the $i$th atom and $\abs{\bm{q}_{j}}=0$ for
all other atoms $j\neq i$, then $\mathrm{PR}=1/N$.  For a plane (sound)
wave of wave-vector $\bm{k}$ we have
$\bm{q}_{i} = \sqrt{2/N}\,\hat{\bm{\zeta}}_{\bm{q}}
\sin\bm{k}\cdot\bm{q}_{i}$
entailing a $\mathrm{PR}=2/3$.  Here $\hat{\bm{\zeta}}_{\bm{q}}$ is
termed the polarization vector and is usually taken as being perpendicular
(transverse sound) or parallel (longitudinal sound) to $\bm{k}$.

\begin{figure}[htb]
  \begin{center}
    \includegraphics[width=.8\linewidth,trim=4cm 4cm 4cm 3cm, clip]{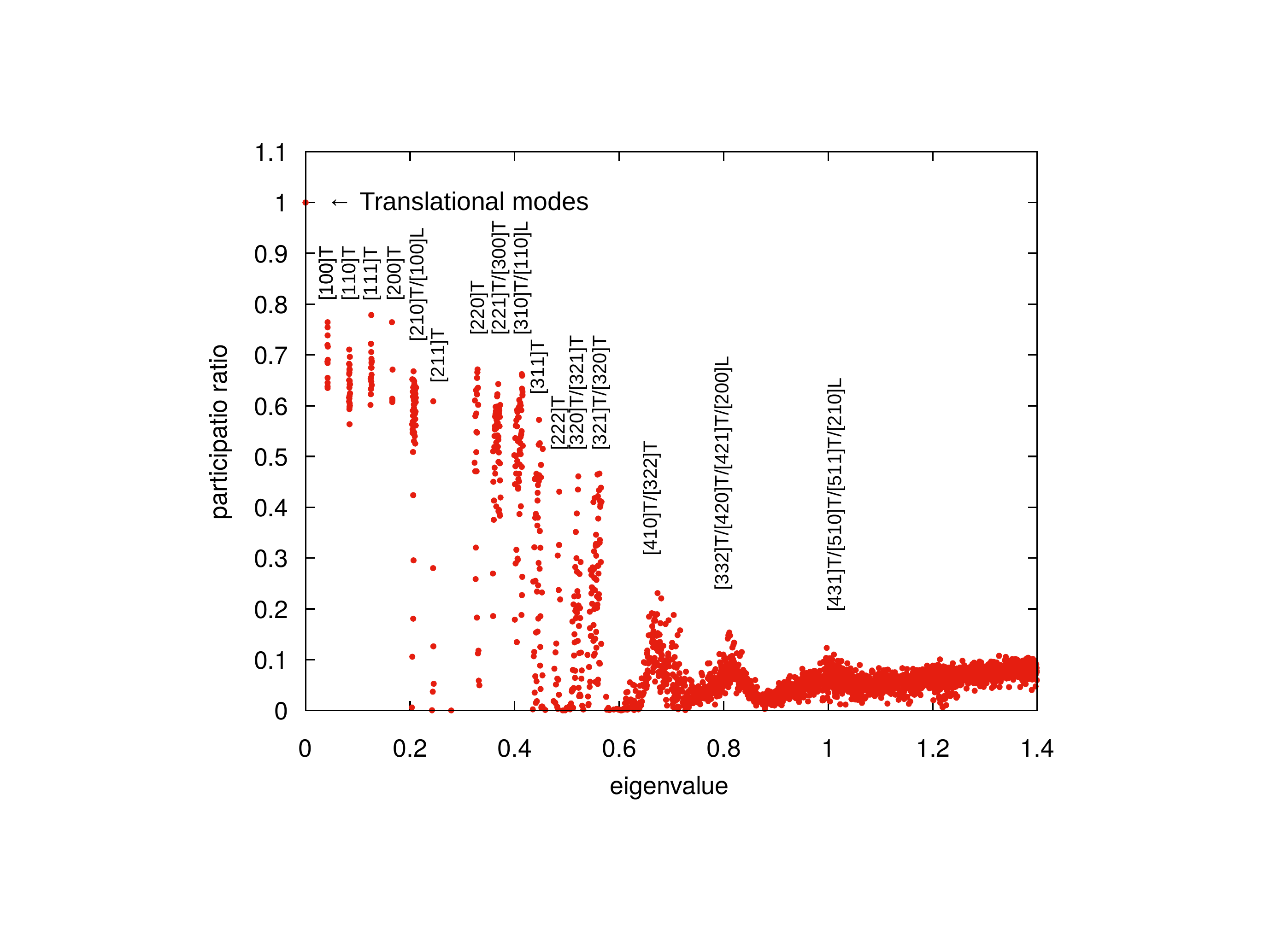}
  \end{center}
  \caption{The participation ratio~\eqref{eq:PR_def} as a function of
    eigenvalue $\lambda$.  At $\lambda=0$, three uniform translational modes
    exist, having a PR equal to unity.  For increasing values of $\lambda$,
    a bunching of eigenvalues is observed, all of which initially have a
    PR=2/3, and correspond to eigenmodes which are well described by the
    plane wave representation
    $[m,n,l]\Leftrightarrow\bm{k}_{mnl}=2\pi/L(m,n,l)$ of either transverse
    (T) or longitudinal (L) polarization.}
  \label{fig:pr}
\end{figure}

Fig.~\ref{fig:pr} plots the participation ratio of the entire
spectrum considered in the present work.  At $\lambda=0$, there
exist three modes with a participation ratio equal to unity, which
correspond to the translational modes of the dynamical matrix.  For the
region up to approximately, $\lambda=0.5$, the eigenvalues are seen to
bunch into clusters with a participation ratio of approximately 2/3,
indicating plane-wave-like eigenmodes and the presence of well defined
sound.  The observed bunching and their multiplicity arise from a
combination of the polarization vectors,
$\hat{\bm{\zeta}_{\bm{k}}}$, and the allowed wave-vectors,
$\bm{k}_{[mnl]}=2\pi/L(m,n,l)$.  Here $L$ is the periodicity length of the
amorphous configuration, and $m$, $n$, and $l$ are integers defining the
allowed wave-lengths.  Through inspection of the corresponding eigenmodes,
a wave-vector family and polarization type can be identified for each
bunching and are indicated in Fig.~\ref{fig:pr}.  As the participation
drops with increasing eigenvalue magnitude, this identification process
becomes more difficult with each peak (now significantly broadened) being
well described by a range of different plane wave components.

\begin{figure}[htb]
  \begin{center}
    \includegraphics[width=1\linewidth,trim=4cm 4cm 4cm 4cm]{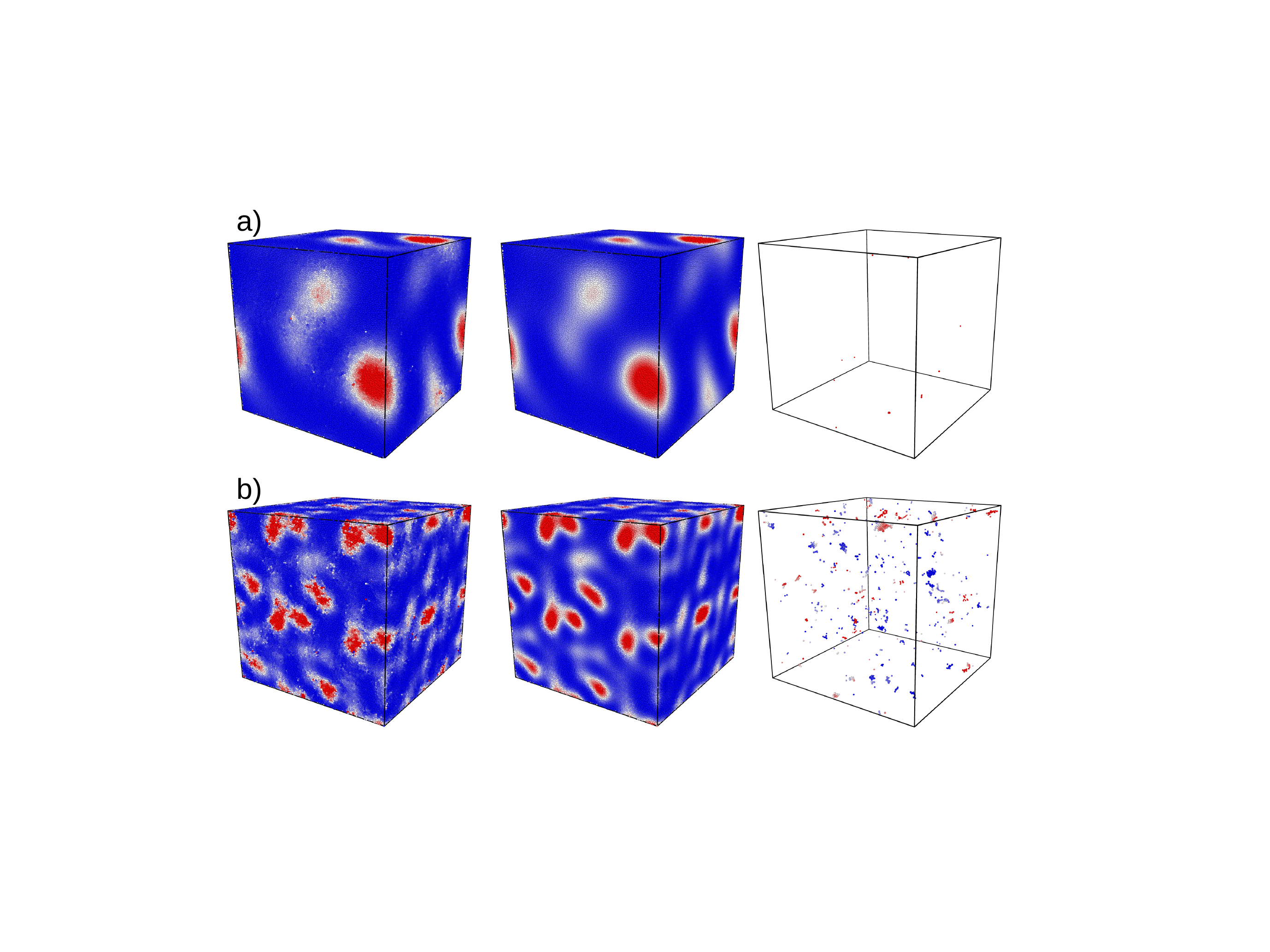}
  \end{center}
  \caption{Plots of the amorphous configuration consisting of 1372000 atoms,
    where each atom is colored according to its value of $|\bm{q}_{i}|^{4}$.
    The left most panels derive its value from the calculated eigenmode and
    central panels from the plane wave decomposition.  The right-most panels
    show only those atoms whose oscillator strength significantly deviates
    from the plane wave decomposition.  a) shows a $[110]$ transverse mode,
    and b) a $[311]$ transverse mode.}
  \label{fig:samples}
\end{figure}

Fig.~\ref{fig:samples} displays the spatial structure of two such
eigenmodes.  In this figure, a) demonstrates a mode that is well described
by $[100]$ transverse plane waves, and b) a mode well described by $[311]$
transverse plane waves.  In both a) and b), three spatial structures are
shown, where the left-most figure plots the atoms at their spatial
coordinates colored according to $|\bm{q}^{}_{i}|^{4}$ derived from the
actual eigenmode and the central figure plots their color according to
$|\bm{q}^{\mathrm{PW}}_{i}|^{4}$ derived from the plane wave (PW)
representation.  The right-most figure plots only those atoms for which
$|\bm{q}_{i}|^{4}>\max\{|\bm{q}^{\mathrm{PW}}_{i}|^{4};i=1,n\}$.
Inspection of the left and central panels of Fig.~\ref{fig:samples}
demonstrates that a large part of the eigenmode derived from the dynamical
matrix is well described by a PW decomposition.  On the other hand, the
right most panels clearly show that their exist local regions of
oscillator strength which are not described by the PW picture.  For the
longer wavelength $[110]$ mode, Fig.~\ref{fig:samples}a, these regions are
rare, but as the wave-length decreases (wave-vector magnitude increases),
such as the $[311]$ mode in Fig.~\ref{fig:samples}b these localized
regions become more numerous and somewhat extended.  For higher
wave-vector magnitudes, this trend continues with an associated drop in
the participation ratio corresponding to the final break up of sound.

Via Fig.~\ref{fig:pr}, both an eigenvalue and wave-vector magnitude regime
can be identified at which the participation ratio rapidly decreases.  The
large system size presently considered allows us to study this regime in
detail, suggesting that a crossover to more heterogeneous modes occurs
over a broad range of eigenvalues.  This corresponds to length scales of
the order of $2\pi/|\bm{k}_{410}|$ to $2\pi/|\bm{k}_{332}|$ and length
scales ranging between 20 and 25 bond lengths.  Such a length-scale is
compatible with amorphous elastic heterogeneity -- a length scale which
is believed to play a defining role in the break up of
sound~\cite{schi:13, msfr:13}.  Larger system sizes will
be needed to investigate whether this cross-over limits to a sharp
transition at a distinct length-scale and particular eigenvalue that may
be finally identified as the boson peak frequency.  It is in such future
simulations, that the true power of the current method will become evident
since all computational resources can now be focused to the actual
eigenvalue region of interest centered around the boson peak frequency.

\section{Conclusions}
\label{sec:concl}

We have discussed a highly parallel implementation of a polynomial filtered
Krylov space-based method for solving large-scale symmetric eigenvalue
problems arising in the investigation of amorphous materials.  The algorithm
enables us to compute hundreds or thousands of eigenvalues of matrices of
size in the millions.  It the number of eigenvalues is too large to compute
at once (e.g.\ for memory requirement) then the interval of interest can be
split in subintervals that contain a reasonable number of eigenvalues whose
associated eigenvectors can be accommodated by the available memory.

The polynomial filter is designed to enhance the eigenvalues of the interval
of interest.  Polynomials of very high degrees result.  Therefore, our
algorithm is based almost completely on matrix-vector multiplications.  The
work to keep the basis vectors of the Krylov space orthogonal is negligible.
This entails a high potential for parallelization which is confirmed by our
experiments.

We plan to apply our solver to problems of size $49'152'000$ and larger.
The use of larger matrices and correspondingly larger simulation cell sizes
will give information on how the vibrational modes of the Boson peak regime
observed in the present work evolve to the bulk limit.  Indeed, the
treatment of larger system sizes will result in a transition to a spectrum
free of eigenvalue bunching, where the effect of disorder smears the allowed
sound-waves into an effectively continuous eigenvalue spectrum.  In this
experimentally relevant limit, the bulk nature of the Boson peak regime
should become manifest from an entirely atomistic description of a model
amorphous system.

We will also work on optimizing the evaluation of matrix
polynomials, in particular, if the matrix polynomials are applied to
multivectors.
As noted in~\eqref{eq:intro.1}, $\bm{A}$ is a matrix formed of
\emph{symmetric} $3\times3$ blocks.  In an efficient implementation of the
eigensolver this should be taken into account.  Doing so,
the arithmetic intensity is more than doubled relative to the standard
elementwise CRS storage format~\cite{TEMPLATES} and execution times can be
reduced by about the same factor.
Finally, let us note that the abundance of matrix-vector multiplications
makes our code amenable to GPU computing.





\appendix
\section{Useful integrals}
\label{sec:integrals}



Here we collect some definite integrals that are needed to compute the
coefficients $\gamma_j$ in~\eqref{eq:cheby-ip.1} with the function $p$ given
in~\eqref{eq:fct_f}.  Note that $p$ is a piecewise quadratic polynomial.
The integral~\eqref{eq:app.1} below has to be evaluated for each of the
three subintervals $[a, b]= [\xi-\varepsilon,\xi]$, $[\xi, \eta]$, and
$[\eta, \eta+\varepsilon]$, individually.

With $1=T_0(x)$, $x=T_1(x)$, $2x^2 = T_2(x)+T_0(x)$, and the
equality~\cite[eq.(22.7.24)]{abst:72}
\begin{equation} \label{eq:app.1}
  2 T_m(x) T_n(x) = T_{m+n}(x) + T_{m-n}(x),
  \qquad m\ge n,
\end{equation}
we get
\begin{align}
  \int_a^b (\alpha + \beta x &+ \gamma x^2) T_j(x)\frac{dx}{\sqrt{1-x^2}} \\[2mm]
  &=   \int_a^b \left[(\alpha + \frac{\gamma}{2})T_0(x) + \beta T_1(x) 
    + \frac{\gamma}{2}T_2(x) \right] T_j(x)\frac{dx}{\sqrt{1-x^2}}\notag \\[2mm]
  &= 
    \begin{cases}
      \displaystyle
      \int_a^b \left[(\alpha + \frac{\gamma}{2}) T_0(x) 
        + \beta  T_1(x)  + \frac{\gamma}{2}T_2(x) \right]
      \frac{dx}{\sqrt{1-x^2}}, & j=0,\notag \\[3mm]
      \displaystyle
      \int_a^b \left[(\alpha + \frac{\gamma}{2}) T_1(x) 
        + \frac{\beta}{2} (T_2(x)+T_0(x))  + \frac{\gamma}{4}(T_3(x)+T_1(x))
      \right] \frac{dx}{\sqrt{1-x^2}}, & j=1, \notag \\[3mm]
      \displaystyle
      \int_a^b \left[(\alpha + \frac{\gamma}{2}) T_j(x) 
        + \frac{\beta}{2} (T_{j+1}(x)+T_{j-1}(x))  +
        \frac{\gamma}{4}(T_{j+2}(x)+T_{j-2}(x)) 
      \right] \frac{dx}{\sqrt{1-x^2}}, & j\ge2.\notag
    \end{cases}
\end{align}
Furthermore, we have
\begin{equation} \label{eq:app.3}
  \int_a^b T_j(x) \frac{dx}{\sqrt{1-x^2}}
  = -\!\!\int\limits_{\arccos(a)}^{\arccos(b)}\!\! \cos j\vartheta\, d\vartheta
  =
  \begin{cases}
    \arccos(a) - \arccos(b), & j = 0, \\[2mm]
    \frac{\displaystyle\sin(j\arccos(a)) - 
      \sin(j\arccos(b))}{\displaystyle j}, & j > 0.
  \end{cases}
\end{equation}



\section*{Acknowledgments}

The computations have been executed on the Euler compute cluster at ETH
Zurich at the expense of a grant of the Seminar for Applied Mathematics.  We
acknowledge the assistance of the Euler Cluster Support Team.

{\small

\begin{thebibliography}{10}

\bibitem{abst:72}
M.~Abramowitz and I.~A. Stegun.
\newblock {\em Handbook of Mathematical Functions}.
\newblock National Bureau of Standards, Washingthon, DC, 10th printing, 1972.
\newblock Available from \url{http://people.maths.ox.ac.uk/~macdonald/aands/}.

\bibitem{acca:17}
G.~Accaputo.
\newblock Solving large scale eigenvalue problems in amorphous materials.
\newblock Master's thesis, ETH Zurich, Computer Science Department, September
  2017.
\newblock \url{doi:10.3929/ethz-b-000221499}.

\bibitem{ahlt:05}
P.~Arbenz, U.~L. Hetmaniuk, R.~B. Lehoucq, and R.~Tuminaro.
\newblock A comparison of eigensolvers for large-scale {3D} modal analysis
  using {AMG}-preconditioned iterative methods.
\newblock {\em Internat. J. Numer. Methods Eng.}, 64(2):204--236, 2005.

\bibitem{avto:11}
H.~Avron and S.~Toledo.
\newblock Randomized algorithms for estimating the trace of an implicit
  symmetric positive semi-definite matrix.
\newblock {\em J. ACM}, 58(2):8:1--8:34, 2011.

\bibitem{bhlt:09}
C.~G. Baker, U.~L. Hetmaniuk, R.~B. Lehoucq, and H.~K. Thornquist.
\newblock {A}nasazi software for the numerical solution of large-scale
  eigenvalue problems.
\newblock {\em ACM Trans. Math. Softw.}, 36(3):1--23, 2009.

\bibitem{bare:16}
M.~{\SortNoop{Barel}}van~Barel.
\newblock Designing rational filter functions for solving eigenvalue problems
  by contour integration.
\newblock {\em Linear Algebra Appl.}, 502:346--365, 2016.

\bibitem{TEMPLATES}
R.~Barrett, M.~Berry, T.~F. Chan, J.~Demmel, J.~Donato, J.~Dongarra,
  V.~Eijkhout, R.~Pozo, C.~Romine, and H.~van~der Vorst.
\newblock {\em Templates for the Solution of Linear Systems: Building Blocks
  for Iterative Methods}.
\newblock SIAM, Philadelphia, PA, 1994.
\newblock (Available from Netlib at URL \url{http://www.netlib.org/templates}).

\bibitem{beks:08}
C.~Bekas, E.~Kokiopoulou, and Y.~Saad.
\newblock Polynomial filtered {L}anczos iterations with applications in density
  functional theory.
\newblock {\em SIAM J. Matrix Anal. Appl.}, 30(1):397--418, 2008.

\bibitem{bede:70}
R.~J. Bell and P.~Dean.
\newblock Atomic vibrations in vitreous silica.
\newblock {\em Discuss. Faraday Soc.}, 50:55--61, 1970.

\bibitem{bcjp:16}
L.~Berthier, P.~Charbonneau, Y.~Jin, G.~Parisi, B.~Seoane, and F.~Zamponi.
\newblock Growing timescales and lengthscales characterizing vibrations of
  amorphous solids.
\newblock {\em Proc. Nat. Acad. Sci.}, 113(30):8397--8401, 2016.

\bibitem{dema:17}
P.~M. Derlet and R.~Maa{\ss}.
\newblock Thermal processing and enthalpy storage of a binary amorphous solid:
  {A} molecular dynamics study.
\newblock {\em J. Mater. Res.}, 32(14):2668--2679, 2017.

\bibitem{dema:18a}
P.~M. Derlet and R.~Maa{\ss}.
\newblock Local volume as a robust structural measure and its connection to
  icosahedral content in a model binary amorphous system.
\newblock {\em Materialia}, 3:97--106, 2018.

\bibitem{dema:18}
P.~M. Derlet and R.~Maa{\ss}.
\newblock Thermally-activated stress relaxation in a model amorphous solid and
  the formation of a system-spanning shear event.
\newblock {\em Acta Mater.}, 143:205--213, 2018.

\bibitem{deml:12}
P.~M. Derlet, R.~Maa{\ss}, and J.~F. L{\"o}ffler.
\newblock The {B}oson peak of model glass systems and its relation to atomic
  structure.
\newblock {\em Eur. Phys. J. B}, 85(5):1--20, 2012.

\bibitem{fasa:12}
H.-R. Fang and Y.~Saad.
\newblock A filtered {L}anczos procedure for extreme and interior eigenvalue
  problems.
\newblock {\em SIAM J. Sci. Comput.}, 34(4):A2220--A2246, 2012.

\bibitem{gkla:17}
M.~Galgon, L.~Kr{\"a}mer, B.~Lang, A.~Alvermann, H.~Fehske, A.~Pieper,
  G.~Hager, M.~Kreutzer, F.~Shahzad, G.~Wellein, A.~Basermann,
  M.~R{\"o}hrig-Z{\"o}llner, and J.~Thies.
\newblock Improved coefficients for polynomial filtering in {ESSEX}.
\newblock In T.~Sakurai, S.-L. Zhang, T.~Imamura, Y.~Yamamoto, Y.~Kuramashi,
  and T.~Hoshi, editors, {\em Eigenvalue Problems: Algorithms, Software and
  Applications in Petascale Computing}, pages 63--79. Springer, 2017.

\bibitem{hutc:90}
M.~F. Hutchinson.
\newblock A stochastic estimator of the trace of the influence matrix for
  {L}aplacian smoothing splines.
\newblock {\em Commun. Stat. -- Simul. Comput.}, 19(2):433--450, 1990.


\bibitem{jksc:99}
L.~O. Jay, H.~Kim, Y.~Saad, and J.~R. Chelikowsky.
\newblock Electronic structure calculations for plane-wave codes without
  diagonalization.
\newblock {\em Comput. Phys. Comm.}, 118(1):21 -- 30, 1999.

\bibitem{kngl:13}
L.~Kr{\"a}mer, E.~Di~Napoli, M.~Galgon, B.~Lang, and P.~Bientinesi.
\newblock Dissecting the {FEAST} algorithm for generalized eigenproblems.
\newblock {\em J. Comput. Appl. Math.}, 244:1--9, 2013.

\bibitem{lanc:56}
C.~Lanczos.
\newblock {\em Applied Analysis}.
\newblock Prentice-Hall, Englewood Cliffs, NJ, 1956.
\newblock (Reprinted by Dover, New York, 1988.).

\bibitem{lxvy:16}
R.~Li, Y.~Xi, E.~Vecharynski, C.~Yang, and Y.~Saad.
\newblock A thick-restart {L}anczos algorithm with polynomial filtering for
  {H}ermitian eigenvalue problems.
\newblock {\em SIAM J. Sci. Comput.}, 38(4):A2512--A2534, 2016.

\bibitem{like:16}
Z.~Liang and P.~Keblinski.
\newblock Sound attenuation in amorphous silica at frequencies near the boson
  peak.
\newblock {\em Phys. Rev. B}, 93(5), 2016.

\bibitem{lisy:16}
L.~Lin, Y.~Saad, and C.~Yang.
\newblock Approximating spectral densities of large matrices.
\newblock {\em SIAM Rev.}, 58(1):34--65, 2016.

\bibitem{msfr:13}
A.~Marruzzo, W.~Schirmacher, A.~Fratalocchi, and G.~Ruocco.
\newblock Heterogeneous shear elasticity of glasses: the origin of the boson
  peak.
\newblock {\em Sci. Rep.}, 3(1407), 2013.

\bibitem{momo:09}
G.~Monaco and S.~Mossa.
\newblock Anomalous properties of the acoustic excitations in glasses on the
  mesoscopic length scale.
\newblock {\em Proc. Nat. Acad. Sci.}, 106(40):16907--16912, 2009.

\bibitem{naps:16}
E.~{\SortNoop{Napoli}}di~Napoli, E.~Polizzi, and Y.~Saad.
\newblock Efficient estimation of eigenvalue counts in an interval.
\newblock {\em Numer. Linear Algebra Appl.}, 23(4):674--692, 2016.

\bibitem{parl:80}
B.~N. Parlett.
\newblock {\em The Symmetric Eigenvalue Problem}.
\newblock Prentice Hall, Englewood Cliffs, NJ, 1980.
\newblock (Republished by SIAM, Philadelphia, 1998.).

\bibitem{rivl:69}
T.~J. Rivlin.
\newblock {\em An Introduction to the Approximation of Functions}.
\newblock Dover, New York, NY, 1981.

\bibitem{ruti:69}
H.~Rutishauser.
\newblock Computational aspects of {F}. {L}. {B}auer's simultaneous iteration
  method.
\newblock {\em Numer. Math.}, 13:4--13, 1969.

\bibitem{saad:11}
Y.~Saad.
\newblock {\em Numerical Methods for Large Eigenvalue Problems}.
\newblock SIAM, Philadelphia, PA, 2nd edition, 2011.
\newblock (Classics in Applied Mathematics; 66).

\bibitem{scha:15}
S.~Schaffner.
\newblock Using {T}rilinos to solve large scale eigenvalue problems in
  amorphous materials.
\newblock Master's thesis, ETH Zurich, Computer Science Department, April 2015.

\bibitem{schi:13}
W.~Schirmacher.
\newblock The boson peak.
\newblock {\em Phys. Status Solidi B}, 250(5):937--943, 2013.

\bibitem{scrs:07}
W.~Schirmacher, G.~Ruocco, and T.~Scopigno.
\newblock Acoustic attenuation in glasses and its relation with the boson peak.
\newblock {\em Phys. Rev. Lett.}, 98(2), 2007.

\bibitem{scsr:15}
W.~Schirmacher, T.~Scopigno, and G.~Ruocco.
\newblock Theory of vibrational anomalies in glasses.
\newblock {\em J. Non-Cryst. Solids}, 407:133--140, 2015.

\bibitem{sccs:12}
G.~Schofield, J.~R. Chelikowsky, and Y.~Saad.
\newblock A spectrum slicing method for the {K}ohn--{S}ham problem.
\newblock {\em Comput. Phys. Comm.}, 183(3):497--505, 2012.

\bibitem{shta:08}
H.~Shintani and H.~Tanaka.
\newblock Universal link between the boson peak and transverse phonons in
  glass.
\newblock {\em Nat. Mat.}, 7(11):870--877, 2008.

\bibitem{siro:96}
R.~N. Silver and H.~{R\"oder}.
\newblock Calculation of densities of states and spectral functions by
  {C}hebyshev recursion and maximum entropy.
\newblock {\em Phys. Rev. E}, 56(4):4822--4829, 1997.

\bibitem{srvk:96}
R.~N. Silver, H.~{R\"oder}, A.~F. Voter, and J.~D. Kress.
\newblock Kernel polynomial approximations for densities of states and spectral
  functions.
\newblock {\em J. Comput. Phys.}, 124(1):115--130, 1996.

\bibitem{sles:03}
G.~L.~G. Sleijpen and J.~{van den Eshof}.
\newblock On the use of harmonic {R}itz pairs in approximating internal
  eigenpairs.
\newblock {\em Linear Algebra Appl.}, 358(1-3):115--137, 2003.

\bibitem{slvo:96}
G.~L.~G. Sleijpen and H.~A. {van der Vorst}.
\newblock A {J}acobi--{D}avidson iteration method for linear eigenvalue
  problems.
\newblock {\em SIAM J. Matrix Anal. Appl.}, 17(2):401--425, 1996.

\bibitem{slvo:00}
G.~L.~G. Sleijpen and H.~A. {van der Vorst}.
\newblock {J}acobi--{D}avidson method.
\newblock In Z.~Bai, J.~Demmel, J.~Dongarra, A.~Ruhe, and H.~van~der Vorst,
  editors, {\em Templates for the solution of Algebraic Eigenvalue Problems: A
  Practical Guide}, pages 238--246. SIAM, Philadelphia, PA, 2000.

\bibitem{tref:13}
L.~N. Trefethen.
\newblock {\em Approximation Theory and Approximation Practice}.
\newblock SIAM, Philadelphia, PA, 2013.

\bibitem{Trilinos-Web-Site}
{The Trilinos Project Home Page}.
\newblock \url{http://trilinos.org/}.

\bibitem{wahn:91}
G.~Wahnstr{\"o}m.
\newblock Molecular-dynamics study of a supercooled 2-component
  {L}ennard--{J}ones system.
\newblock {\em Phys. Rev. A}, 44(6):3752--3764, 1991.

\bibitem{wwaf:06}
A.~Wei{\ss}e, G.~Wellein, A.~Alvermann, and H.~Fehske.
\newblock The kernel polynomial method.
\newblock {\em Rev. Mod. Phys.}, 78:275--306, 2006.

\bibitem{wusi:00}
K.~Wu and H.~D. Simon.
\newblock Thick-restart {L}anczos method for large symmetric eigenvalue
  problems.
\newblock {\em SIAM J. Matrix Anal. Appl.}, 22(2):602--616, 2000.

\bibitem{xwln:07}
N.~Xu, M.~Wyart, A.~J. Liu, and S.~R. Nagel.
\newblock Excess vibrational modes and the {B}oson peak in model glasses.
\newblock {\em Phys. Rev. Lett.}, 98(17), 2007.

\bibitem{ytsi:12}
I.~Yamazaki, H.~Tadano, T.~Sakurai, and T.~Ikegami.
\newblock Performance comparison of parallel eigensolvers based on a contour
  integral method and a {L}anczos method.
\newblock {\em Parallel Comput.}, 39(6):280--290, 2013.

\bibitem{zstc:06}
Y.~Zhou, Y.~Saad, M.~L. Tiago, and J.~R. Chelikowsky.
\newblock Self-consistent field calculations using {C}hebyshev-filtered
  subspace iteration.
\newblock {\em J. Comput. Phys.}, 219(1):172--184, 2006.

\end{thebibliography}
\providecommand{\SortNoop}[1]{}

}

\end{document}